\documentclass[utf8]{FrontiersinHarvard}
\usepackage{url,hyperref,lineno,microtype}
\usepackage[onehalfspacing]{setspace}

\usepackage{graphicx}
\usepackage{subcaption}
\usepackage{siunitx}

\newcommand{\vlos}{$v_{\mathrm{LOS}}$~}
\newcommand{\blos}{$B_{\mathrm{LOS}}$~}

\def\keyFont{\fontsize{8}{11}\helveticabold }
\def\firstAuthorLast{Skirvin {et~al.}} 
\def\Authors{Samuel J. Skirvin\,$^{1,2,*}$, Samuel D. T. Grant\,$^{2}$, David B. Jess\,$^{2,3}$, Ryan J. Campbell\,$^{2}$, Shahin Jafarzadeh\,$^{2}$, Mika V. Kontiainen\,$^{4,2}$, Michele Berretti\,$^{5,6}$, Timothy J. Duckenfield\,$^{2}$, Glen Chambers\,$^{2}$, Marco Stangalini\,$^{7}$ and Luc Rouppe van der Voort\,$^{8,9}$}

\def\Address{$^{1}$School of Engineering, Physics and Mathematics, Northumbria University, Newcastle Upon Tyne, NE1 8ST, UK \\
$^{2}$Astrophysics Research Centre, School of Mathematics and Physics, Queen’s University Belfast, Belfast, BT7 1NN, UK \\
$^{3}$Department of Physics and Astronomy, California State University Northridge, Northridge, CA 91330, USA \\
$^{4}$Institute of Astronomy, University of Cambridge, Madingley Road, CB3 0HA, UK \\
$^{5}$University of Trento, Via Calepina 14, 38122 Trento, Italy \\
$^{6}$University of Rome Tor Vergata, Department of Physics, Via della Ricerca Scientifica 3, 00133 Rome, Italy \\ 
$^{7}$ASI, Italian Space Agency, Via del Politecnico snc, 00133, Rome, Italy \\
$^{8}$Institute of Theoretical Astrophysics, University of Oslo, Oslo, Norway \\
$^{9}$Rosseland Centre for Solar Physics, University of Oslo, Oslo, Norway}

\def\corrAuthor{Samuel J. Skirvin}

\def\corrEmail{samuel.j.skirvin@northumbria.ac.uk}

\begin{document}
\onecolumn
\firstpage{1}

\title{The Influence of Opacity on Inferred MHD Wave Signatures in the Lower Solar Atmosphere} 

\author[\firstAuthorLast ]{\Authors} 
\address{} 
\correspondance{} 

\extraAuth{}

\maketitle

\begin{abstract}

Magnetohydrodynamic wave activity in small-scale magnetic structures, such as solar pores, provides key insights into energy transport in the lower solar atmosphere. This study presents high-resolution observations of ten solar pores contained within a $43 \times 43$~Mm$^{2}$ field of view, obtained with the CRISP instrument at the Swedish 1-m Solar Telescope. We investigate the temporal behaviour of the line-of-sight velocity (\vlos) and magnetic field (\blos) oscillations within the pore structures. Using SIR inversions, we analyse the oscillatory signals at multiple optical depths ($\log \tau$ levels) to assess how variations in geometric height ($z$) and temperature relate to the observed \blos fluctuations. Our results reveal that higher-frequency oscillations ($>6$~mHz) exhibit strong coherences with in-phase fluctuations between \blos and $z$ across atmospheric layers, consistent with upward-propagating magneto-acoustic waves. In contrast, coherent lower-frequency oscillations display significant phase differences, which may arise from opacity effects contaminating the inversion response. These findings highlight the importance of accounting for opacity effects when interpreting magnetic oscillations, with direct implications for forthcoming high-precision magnetic diagnostics from facilities such as DKIST.

\tiny
 \keyFont{ \section{Keywords:} Solar photosphere, Magnetohydrodynamics, Solar oscillations, Solar active region magnetic fields, MHD Waves} 
\end{abstract}

\section{Introduction}

The solar atmosphere is replete with oscillations interpreted as the presence of magnetohydrodynamic (MHD) waves \citep{Jess2015, VD2020, Jess2023, Morton2023}, which are important for the dynamical coupling, energy transport and potential plasma heating across the solar atmosphere. Solar pores, regions of stronger magnetic field often related to the formation or breaking of a sunspot \citep{Garcia1987}, typically possess mean field strengths of $1200 - 1600$~G and mean inclinations of $30^{\circ}$  \citep{Sob2003, Rozo2023}. Their smaller size compared to sunspots permits a greater influence from localised drivers such as granular buffeting \citep{Riedl2021}, resulting in the ubiquitous generation of global MHD wave modes. Solar pores have been shown to support various MHD modes such as sausage modes \citep{Doro2008, Morton2011, Grant2015, Keys2018, GM21, Grant2022}, in addition to higher order kink modes and fluting modes \citep{Jaf2024}.

Capturing signatures of magnetic field oscillations has been of great interest for decades, however, two primary factors have prohibited detailed study; the instrumental sensitivity required to detect incremental polarimetric signatures \citep{Bailen2023} and the need to develop robust inversion techniques to extract the magnetic field information \citep{Rod2017}. As advancements have been made on both fronts, observational campaigns have begun to reveal the oscillatory nature of the strong magnetic fields in the lower atmosphere \citep{Fujimura2009,Houston2018, Joshi2018, Nelson2021, Stan2021, Murabito2021}. These works were able to recover behaviours similar to those seen in velocity data, such as the predominance of the $p$-mode spectrum, although some discrepancies seemed to indicate that these atmospheric parameters were not mirroring one another. For example, \citet{Stangalini2021} provided the first indication of a disconnect between velocity and magnetic field. A comparison between circular polarisation (deployed as a proxy of magnetic field) and velocity revealed distinctive power spectra in a magnetic pore. Their peak-frequencies were offset, with line-of-sight (LOS) velocity showing increased higher frequency power. The implications of magnetic and velocity fields displaying distinct wave energetics are vast, implying that magnetic oscillations could hold novel truths about the solar atmosphere. It is therefore vital to study this disconnect in greater detail.

A recent analysis by \citet{Felipe2023} explored the extent to which NLTE inversions accurately recover chromospheric oscillations produced by wave propagation, finding that height-dependent opacity effects significantly modify the retrieved velocity and temperature oscillations and become especially important for frequencies exceeding $9$~mHz. Moreover, a recent observational study by \citet{Grant2022} examined the propagation of coherent magnetohydrodynamic waves across multiple solar magnetic pores, finding ubiquitous slow sausage-mode oscillations in the photosphere that remain coherent at low heights but become decorrelated in the chromosphere, with evidence of power amplification near $5$~mHz and complex height-dependent wave morphology.

In this study, we examine high-resolution data obtained from the Swedish Solar Telescope to investigate the influence of opacity, if any, on the observed solar $p$-mode spectrum and search for further evidence of MHD wave propagation across a number of solar pores. The paper is structured as follows: in Section \ref{sec:data} we outline the observations used in the current work, and the procedure of identifying the pores and their properties. Section \ref{sec:results} highlights the results of the wave analysis on the SIR inversions. Finally, in Section \ref{sec:conclusions} we provide a brief summary of our results, implications for interpreting oscillations in the solar photosphere and necessary avenues of future work. 

\section{Observations and Data Processing}\label{sec:data}

\begin{figure*}[t!]
  \centering
  \includegraphics[trim=0mm 0mm 0mm 0mm, clip, width=0.99\textwidth, angle=0]{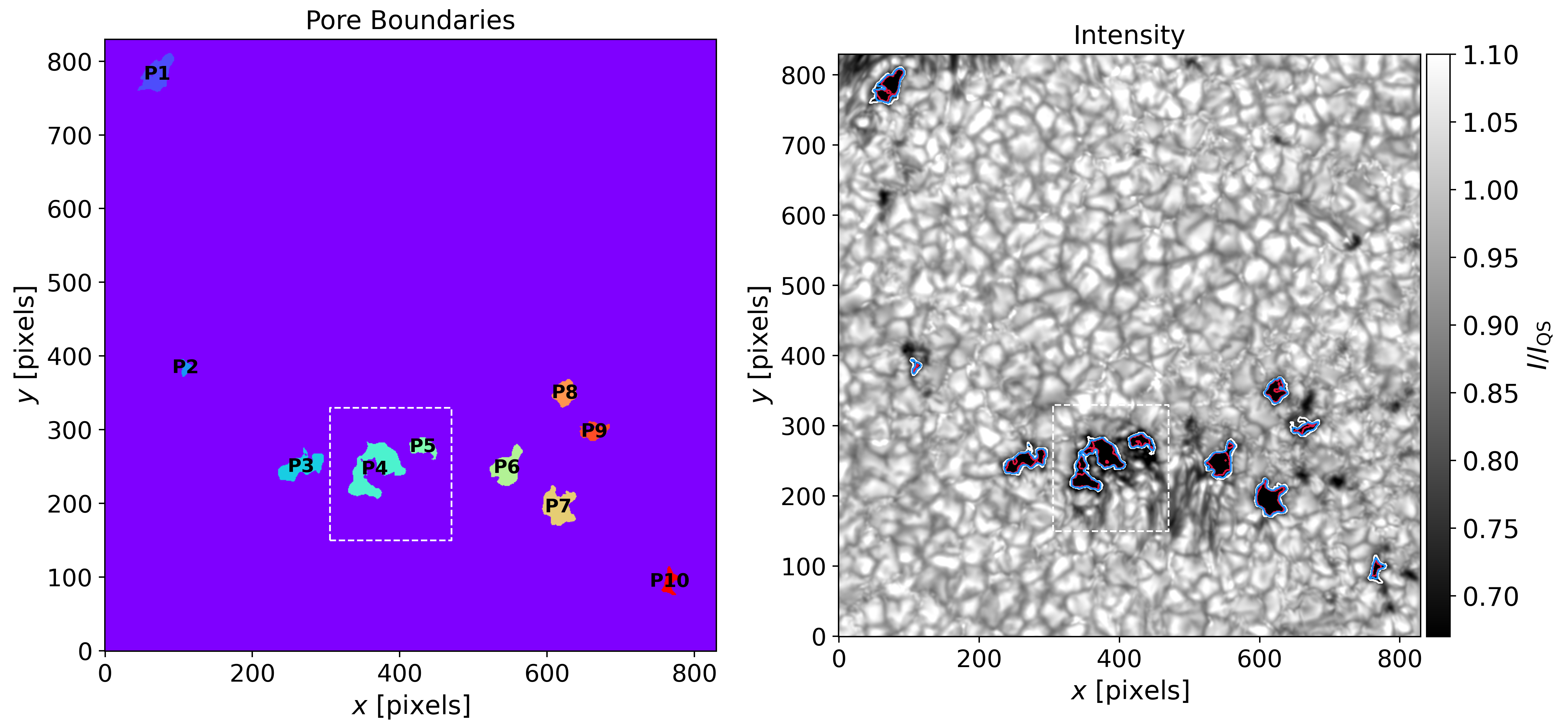}
   \includegraphics[trim=0mm 0mm 0mm 0mm, clip, width=0.97\textwidth, angle=0]{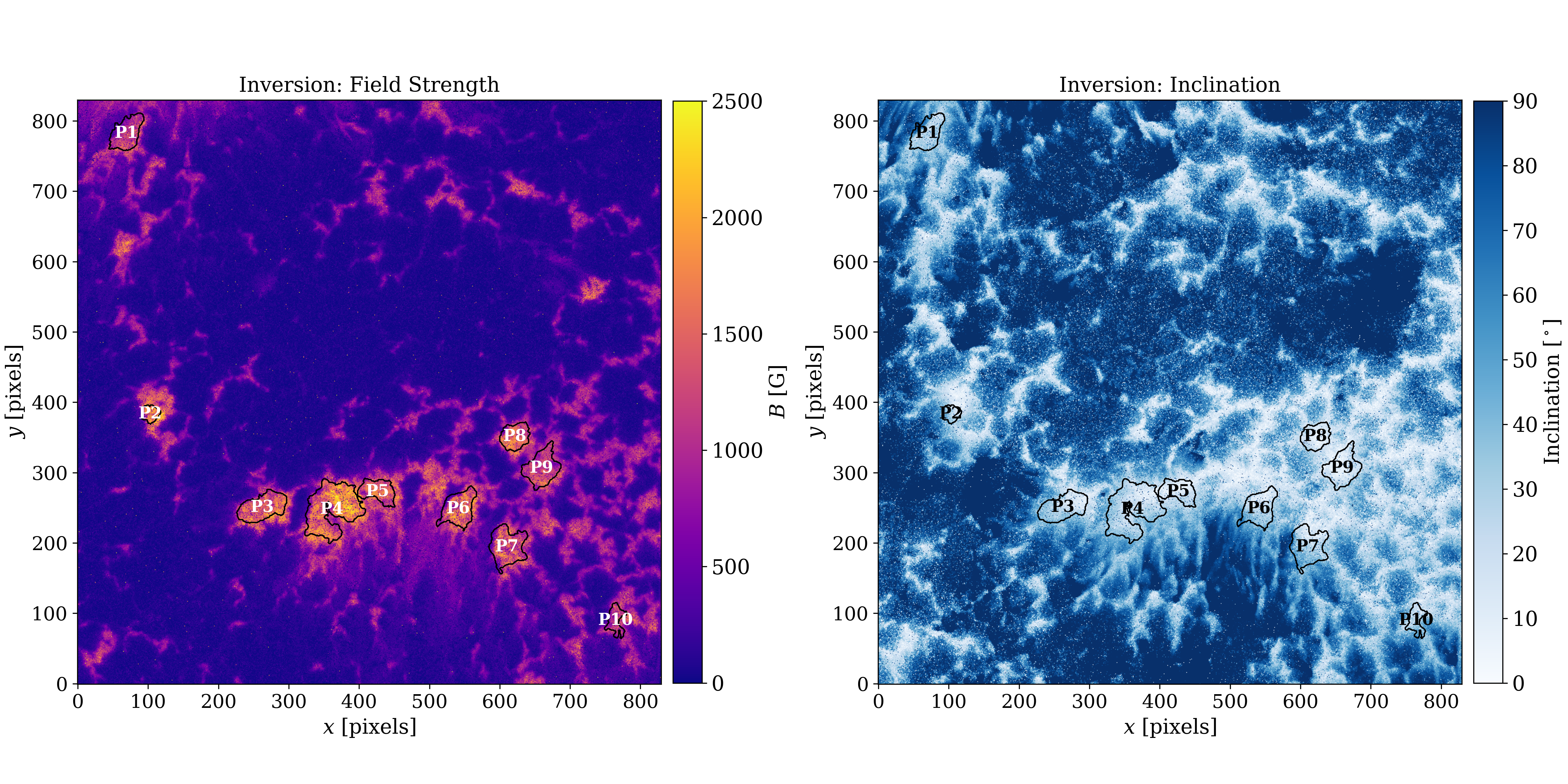}
\caption{Reduced field of view for the region of interest. (Top Left): Labelled maximum pore boundaries P1--P10. (Top Right): Mean continuum intensity with regions identified as belonging to a pore at any point in the time series (white), in the mean (blue), and throughout the time series (red). The dashed white box is the active region within which a different intensity threshold is used to minimise interference from the emerging penumbral region and spurious coalescence of the neighbouring pores. (Bottom left): Total magnetic field strength obtained from the SIR inversion. (Bottom right): Inclination angle at the initial time of observation from the SIR inversion.}
\label{Fig: FoV}
\end{figure*}

The data presented here is an observational sequence obtained with the 1-m Swedish Solar Telescope \citep[SST;][]{Scharmer2003, Scharmer2019} during $08$:$52-09$:$45$~UT on 2016 September 05. The region of interest was active region NOAA $12585$, with the field-of-view (FoV) centred around a group of magnetic pores at heliocentric co-ordinates (\SI{-172}{\arcsecond}, \SI{22}{\arcsecond}), corresponding to the cosine of an observing angle of $\mu = 0.98$. The CRisp Imaging SPectropolarimeter \citep[CRISP;][]{Scharmer2008} was utilised to provide spectro-polarimetric imaging data of the Fe {\sc{i}} $6301$ \& $6302$~\AA~line pair sampling 16 non-equidistant wavelength steps from $-1.17$ - $0.08$~\AA~around the Fe {\sc{i}} $6302$~\AA~line-core. The CRISP instrument imaged a \SI{59}{\arcsecond} $\times$ \SI{59}{\arcsecond} (approximately $43 \times 43$~Mm$^{2}$) region of the solar disk (see Figure~\ref{Fig: FoV}), with a spatial sampling of \SI{0.059}{\arcsecond} (43{\,}km) per pixel and temporal cadence of 32.2{\,}s for a total of 98 scans. Seeing remained good throughout, with data fidelity improved through the use of high-order adaptive optics \citep{Scharmer2003b}, alongside processing with the CRISPRED pipeline \citep{Rodriguez2015}, Multi-Object Multi-Frame Blind Deconvolution \citep[MOMFBD;][]{Noort2005} image restoration and image destretching. 

The data was inverted using the Stokes Inversion based on Response functions \citep[SIR;][]{Ruiz1992} code under the assumption of local thermodynamic equilibrium (LTE) in order to infer the physical parameters of the atmosphere for the full field-of-view. Each Stokes vector was inverted multiple times using randomised initial values for magnetic field strength, $B$, line-of-sight velocity, $v_{\mathrm{LOS}}$, magnetic inclination, $\gamma$, and magnetic azimuth, $\phi$, with only the best-fit solution retained. This reduces sensitivity to initial conditions and reduces the probability of convergence to local, as opposed to global, minima. Elemental abundances were adopted from \citet{asplund}. One inversion cycle was employed with five nodes in temperature, three nodes each in $B$ and $v_{\mathrm{LOS}}$, two in $\gamma$, and one in $\phi$.

We compute the inversions at two different optical depths corresponding to $\log\tau=-1$ and $\log\tau=-2$. This is because the Fe~{\sc{i}} $6301$~{\AA} and $6302$~{\AA} line pair Stokes~$V$ sensitivity to \blos is broad, but typically peaks within the range $\log\tau = [–1.3:-1.7]$ \citep{Quintero2021}. Therefore, these optical depths are representative of the most accurately inferred \blos signals and are utilised to provide a physical picture of where the \blos signal originates in pores observed in the SST FoV. 

\subsection{Pore Identification}
Solar pores are regions of strong magnetic field in the Sun's photosphere, hence they appear dark in intensity maps due to suppressed plasma convection. At each time step, the continuum intensity was normalised with respect to the mean intensity of the quiescent Sun, $I_{\rm QS}$, defined as sub-field where the LOS magnetic field strength was $<50$~G. The normalised frames were combined to construct a mean continuum intensity frame (Figure \ref{Fig: FoV}). The maximum boundaries for pore candidates (white contours) were then defined as contiguous groupings of pixels that at some point in the time series satisfy $I/I_{\rm QS} \leq 0.6$ in the active region (dashed white box) and $I/I_{\rm QS} \leq 0.7$ elsewhere. To ensure their suitability for wavelet analysis, only those pore candidates were retained for which a corresponding structure greater than 100 pixels in area could be identified in the mean continuum intensity frame (blue contours). The resulting pore boundaries were labelled P1--P10 from left to right, and the pores in each frame of time series were identified as groupings of pixels satisfying the intensity thresholds within the established maximum pore boundaries. To rule out potential false signals due to incoherent oscillations within the pores, the pores with non-vanishing permanent areas (1, 3, 4, 6, 7, 8, 9) were chosen for further analysis and are discussed in the following sections. The three excluded pores (P2, P5, P10) were omitted as they did not maintain coherent structures which were large enough to be confidently traced throughout the duration of the observation. For example, P5 temporarily merges with P4, while P10 appears only intermittently above the intensity threshold. P2, the largest of the excluded pores, was found to exhibit an energy spectral density comparable to the analysed pores (e.g. P8, displayed in Figure \ref{Pore_8_InterPore_ESD}), suggesting the selection criterion does not introduce a systematic bias in the observed wave statistics.

\subsection{Pore Properties}

\begin{table*}[ht]
\caption{Average properties of the seven pores identified at the solar surface and used for further analysis. The pore area, \vlos, magnetic field strength, magnetic field inclination and temperature were calculated from the SIR inversions at log$(\tau)=-1$. Note that magnetic field is directed into the Sun (i.e. \blos $< 0$) so the inclination angles have been calculated using `$180 - $ inclination'.}
\label{Pore Properties}      
 \centering
\begin{tabular}{l c c c c c}
\hline\hline
Pore & {Area (Mm$^{2}$)} & {LoS Velocity (km s$^{-1}$)} & {Magnetic Field (G)} & {Inclination ($^{\circ}$)} & Temperature ($K$) \\    
\hline                        
Pore 1 & $1.48 \pm 0.40$ & $0.47 \pm 0.46$ & $1137 \pm 225$ & $28.0 \pm 7.6$ & $4786 \pm 105$ \\ 
Pore 3  & $2.17 \pm 0.31$ & $0.57 \pm 0.46$ & $1407 \pm 373$ & $24.4 \pm 10.8$ & $4744 \pm 114$ \\
Pore 4  & $4.06 \pm 0.43$ & $0.44 \pm 0.33$ & $1812 \pm 465$ & $20.2 \pm 19.3$ & $4552 \pm 97$ \\
Pore 6  & $1.76 \pm 0.33$ & $0.52 \pm 0.47$ & $1336 \pm 396$ & $21.2 \pm 15.1$ & $4766 \pm 124$ \\
Pore 7  & $2.26 \pm 0.22$ & $0.56 \pm 0.53$ & $1218 \pm 314$ & $25.4 \pm 11.2$ & $4579 \pm 80$ \\
Pore 8  & $1.31 \pm 0.13$ & $0.69 \pm 0.74$ & $1430 \pm 397$ & $17.5 \pm 28.4$ & $4813 \pm 97$ \\
Pore 9  & $1.38 \pm 0.35$ & $0.73 \pm 0.67$ & $1219 \pm 307$ & $15.9 \pm 18.2$ & $4774 \pm 154$ \\
\hline
\end{tabular}
\end{table*}

To determine the pore morphology, we utilized the binary map of the maximum pore boundaries where the barycentre of each individual pore was found using \texttt{scipy.ndimage.center\_of\_mass} and both barycentric minimum circumscribed and maximum inscribed circles fitted to the pore pixels \citep{SUI20122159}. After conducting the SIR inversions, we can use the permanent pore pixels to determine the average line of sight velocity, magnetic field strength, magnetic field inclination and plasma temperature within each pore, where the values are outlined in Table \ref{Pore Properties}.

\section{Analysis and Discussion}\label{sec:results}

In this work, we wish to study whether opacity or geometric effects can contribute to the apparent coherence between \vlos and \blos signals in a group of solar pores, and may be incorrectly interpreted as real magnetic oscillations (i.e. MHD waves). Moreover, we wish to understand the effect, if any, of opacity effects modifying the observed power spectrum.

\subsection{Wave Activity}

\begin{figure}[t!]
  \centering
  \includegraphics[trim=0mm 0mm 0mm 0mm, clip, width=0.8\textwidth, angle=0]{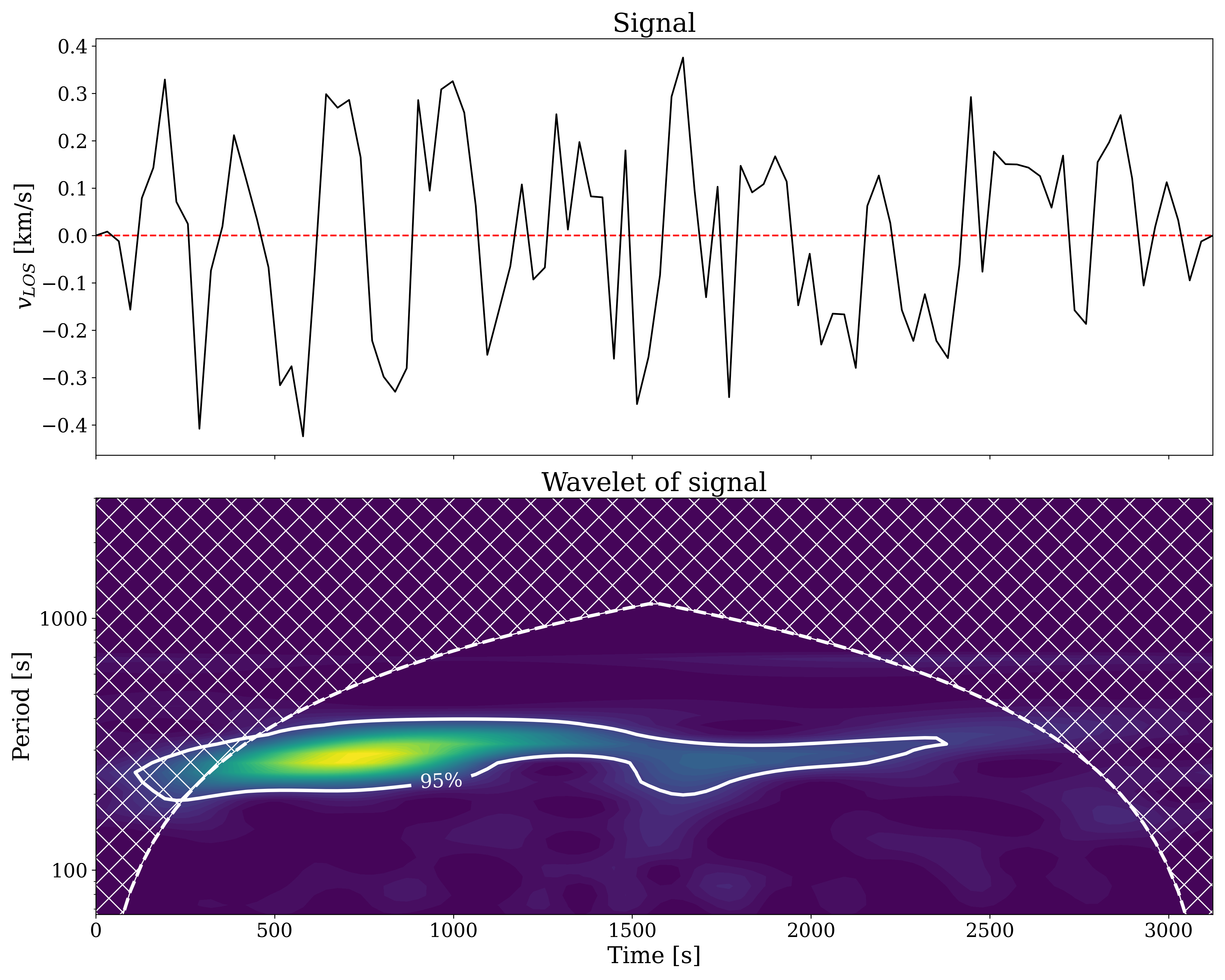}
\caption{Wavelet power spectrum of \vlos time series for P8. Prior to the wavelet transform, the time series was detrended with a 3rd-order polynomial and apodised with a 90\% tapered cosine filter.}
\label{Pore_8_Wavelet}
\end{figure}

In order to analyse wave activity in the identified pores, the \vlos and \blos time series for each pore with non-vanishing permanent areas (e.g. pores 1, 3, 4, 6, 7, 8, 9), were obtained using the binary maps, detrended with a linear fit, and apodised using a Tukey window with taper fraction $\alpha$ = 0.1. This yielded appropriate time series for Morlet wavelet analysis, an example of which is shown in Figure \ref{Pore_8_Wavelet} for the \vlos signal of pore 8. The duration of wave activity at 95\% significance for each frequency/period was calculated as a fraction of the total length of the time series. Statistical significance was assessed against a red noise (AR(1)) background spectrum using the method of \citet{Torrence1998}, where the lag-1 autocorrelation was estimated from the data. We interpret only periods for which at least three oscillation cycles are supported outside the cone-of-influence (COI). For a Morlet wavelet, the COI e-folding time scales as $\sqrt{2}\,s_{\rm w}$, which motivates excluding periods approaching the total duration $T$ of the time series; in practice we adopt an effective maximum reliable period of order $P_{\max}\sim T/(3\sqrt{2})$.

\begin{figure}[t!]
  \centering
  \includegraphics[width=0.9\linewidth]{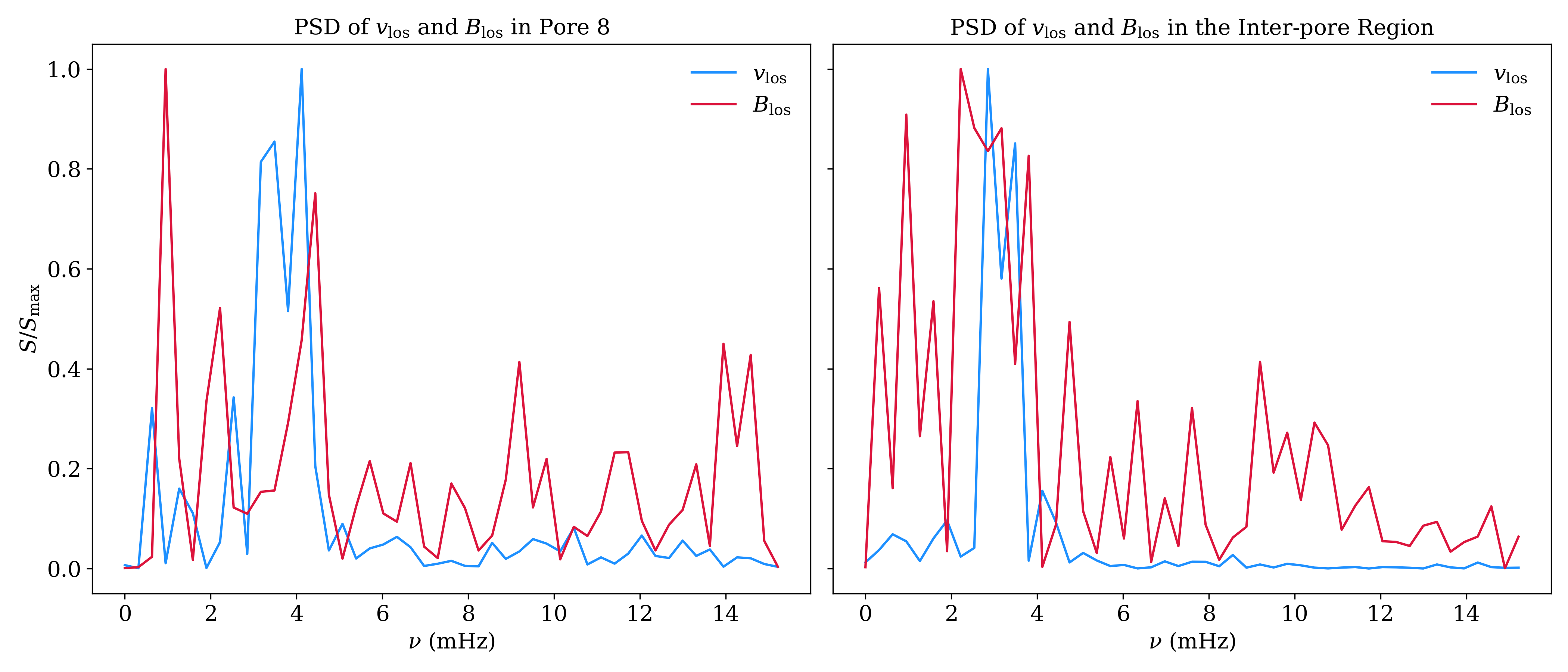}

\caption{Normalised power spectral density as calculated from the detrended and apodised \vlos (blue) and \blos (red) time series, determined from the SIR inversion, for pore~$8$ (left) and the inter-pore region (right) identified as being free of pores for the duration of the time series.}
\label{Pore_8_InterPore_ESD}
\end{figure}

\begin{figure}[t!]
  \centering
  \includegraphics[width=0.6\linewidth]{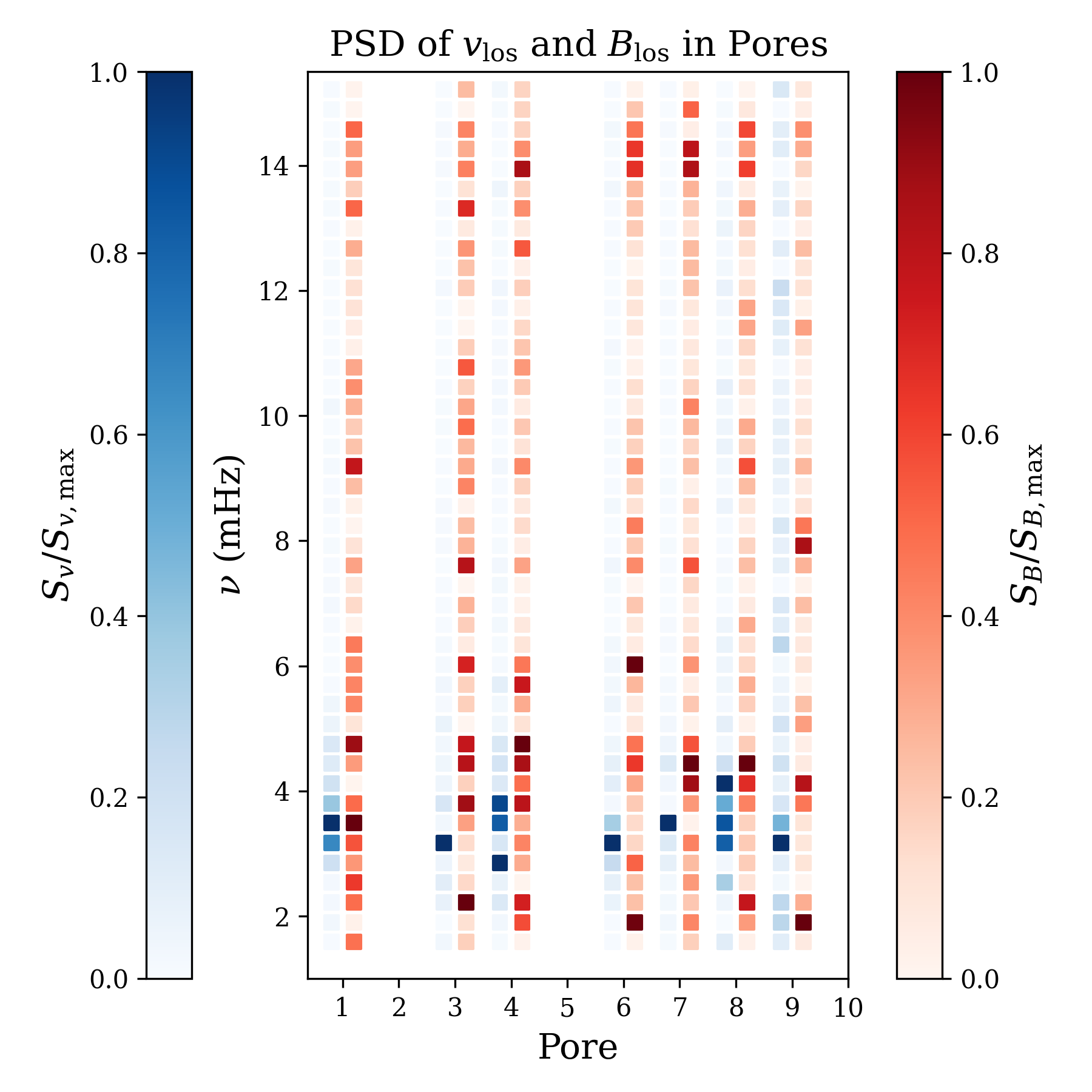}
\caption{Normalised power spectral density as calculated from the detrended and apodised \vlos (blue) and \blos (red) time series for each of the pores. In the former, the $3$~mHz photospheric oscillations dominate the PSD.}
\label{Pores_ESD}
\end{figure}

Using the detrended and apodised \vlos and \blos time series, the power spectral density (PSD) for each pore was obtained following \citet{Grant2022}. We performed a fast Fourier transform on the detrended time series. The resulting Fourier transform $X(f)$ (complex amplitude) was used to calculate the power spectral density given by:
\begin{equation}
S(f) = \frac{2|X(f)|^2}{\delta \nu},
\end{equation}
where $S$ represents the PSD and $\delta \nu$ is the frequency sampling. As an example, Figure \ref{Pore_8_InterPore_ESD} displays the resulting PSD for P8 and the inter-pore region, while the distribution for all pores is shown in Figure \ref{Pores_ESD}. It is clear that the \vlos signal in both the pore and interpore regions, displays a strong power around $3$~mHz. The \blos signal in the interpore region also peaks around a similar frequency, however, within the pore, there is significant power at both lower and higher frequencies for \blos.

\subsection{Phase Differences}
To determine whether the wave activity seen across the pores arises from the same physical source, the phase differences between LOS velocity, LOS magnetic field, geometrical height and temperature oscillations across all pores were calculated in a pair-wise fashion using a wavelet cross-correlation approach. 

\setcounter{figure}{5}
\setcounter{subfigure}{0}
\begin{subfigure}
\setcounter{figure}{5}
\setcounter{subfigure}{0}
    \centering
    \begin{minipage}[b]{0.48\textwidth}
        \includegraphics[width=\linewidth]{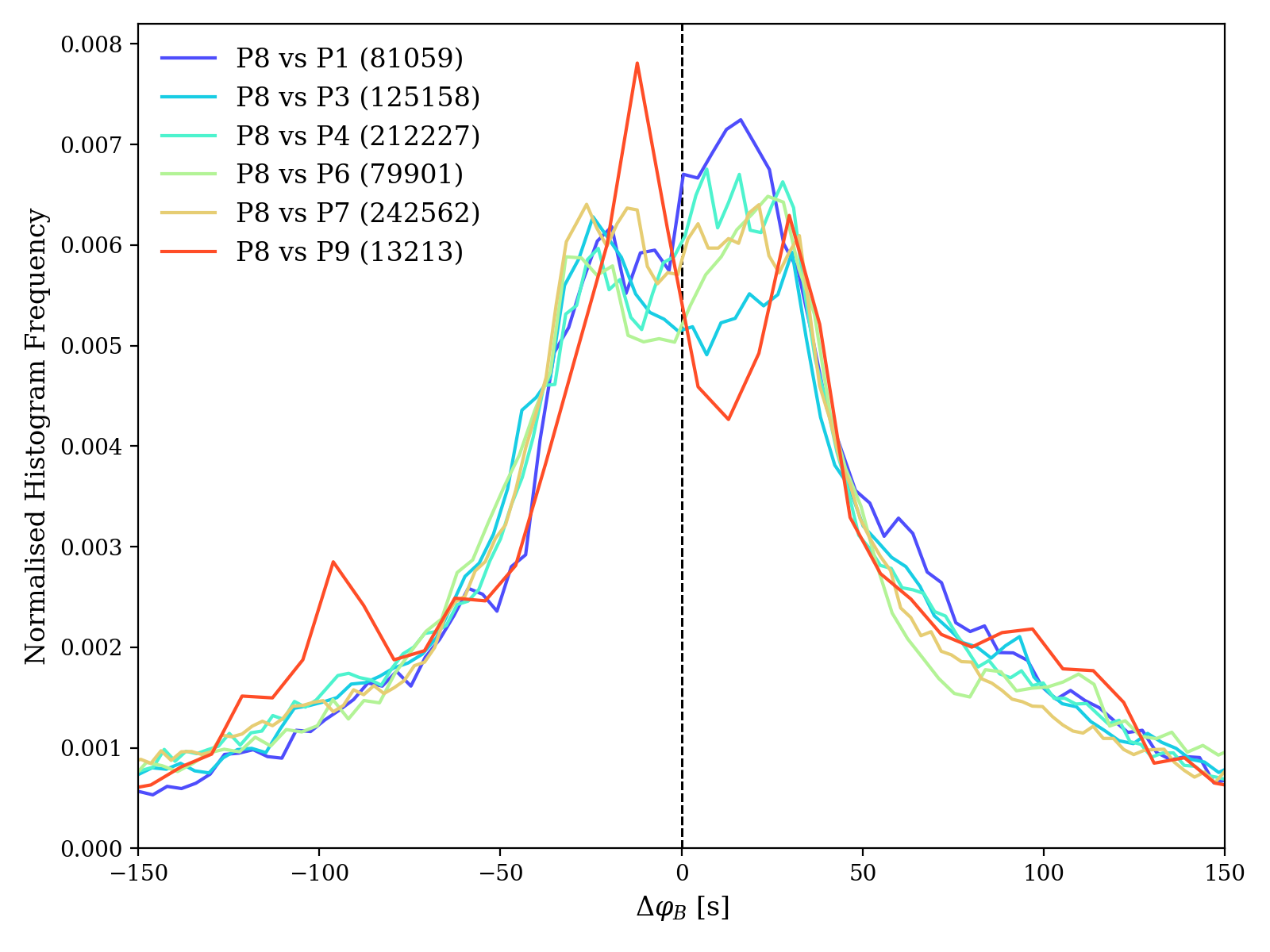}
        \caption{}  
        \label{fig:p8_vs_phase_blos}
    \end{minipage}  
\hfill   
\setcounter{figure}{5}
\setcounter{subfigure}{1}
    \begin{minipage}[b]{0.48\textwidth}
        \includegraphics[width=\linewidth]{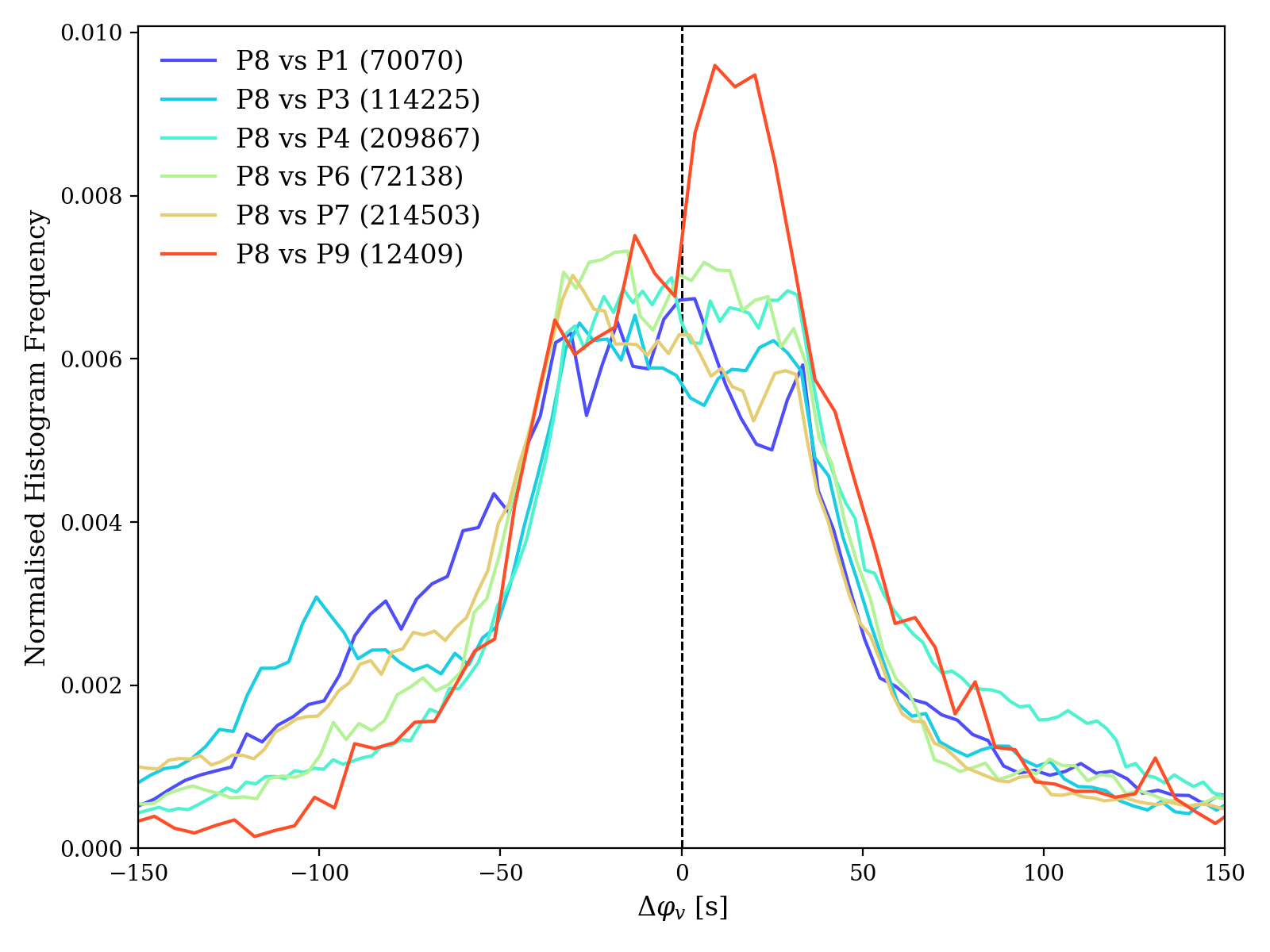}
        \caption{}  
        \label{fig:p8_vs_phase_vlos}
    \end{minipage}
\setcounter{figure}{5}
\setcounter{subfigure}{-1}
    \caption{Distribution of phase differences, displayed as time lags, of the (a) \blos and (b) \vlos signals from pore 8 against the other pores considered in the analysis for log$(\tau) = -1$.}
    \label{fig:VLOS_BLOS_WAVELET_p8VSall}
\end{subfigure}

To rule out potential false signals due to incoherent oscillations within the pores, the pores with non-vanishing permanent areas were chosen for further pixel-by-pixel analysis. Phase differences in coherent velocity and magnetic field oscillations between the barycentric pixel and all other constant pixels were obtained using a Morlet coherence-wavelet transform with the same parameters as outlined above, with period greater than $T/3\sqrt{2}$ neglected. The resulting histograms for the phase lags within the pores (Figure \ref{fig:VLOS_BLOS_WAVELET_p8VSall}) display a largely normal distribution and are roughly concentrated on zero, for both \blos and \vlos signals, indicating that the oscillations within pores are largely coherent. However, both distributions are quite broad, which may suggest that the signals, potentially containing real magnetic oscillations, are being contaminated by a separate source.

\setcounter{figure}{6}
\setcounter{subfigure}{0}
\begin{subfigure}
\setcounter{figure}{6}
\setcounter{subfigure}{0}
    \centering
    \begin{minipage}[b]{0.48\textwidth}
        \includegraphics[width=\linewidth]{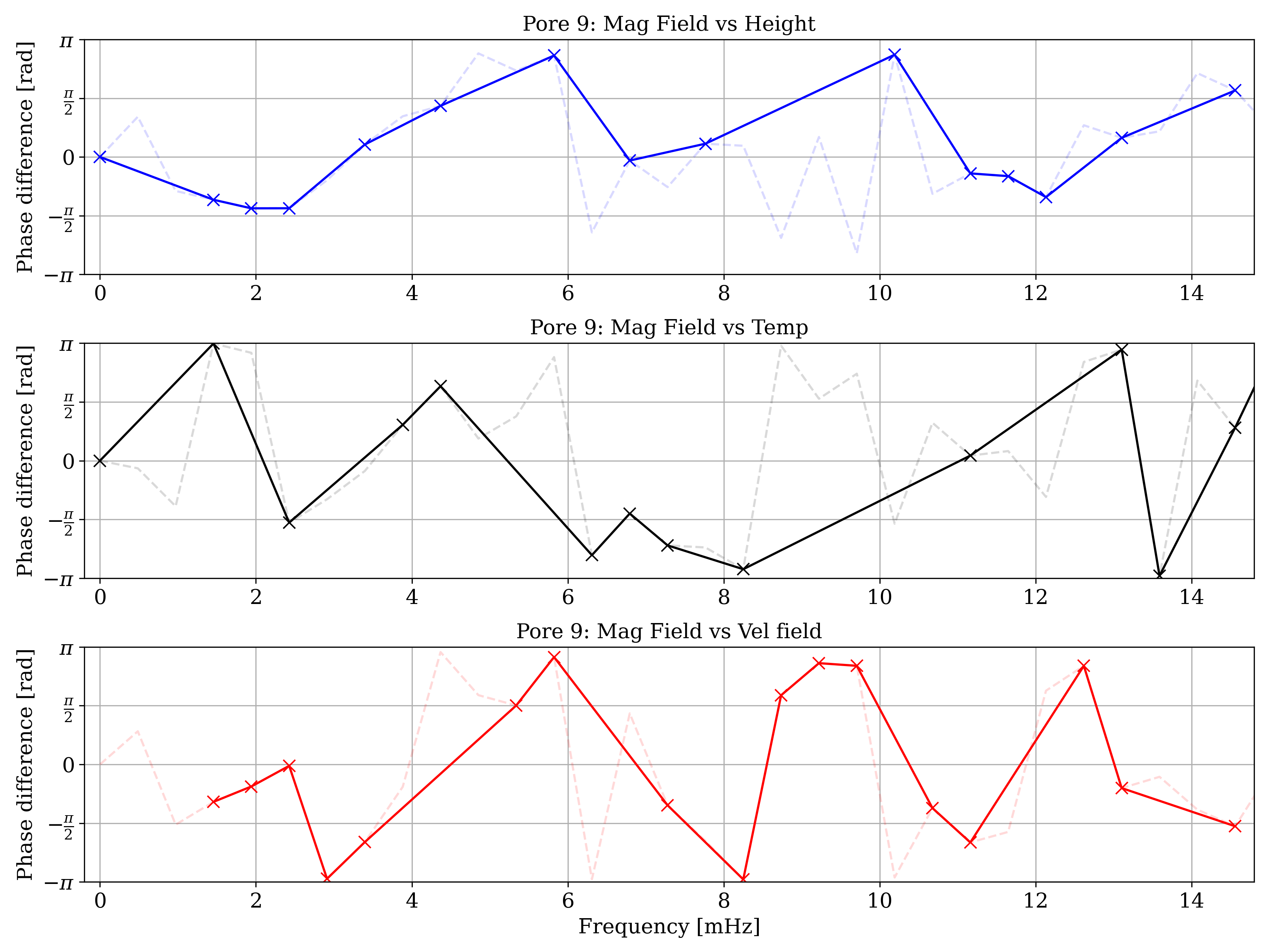}
        \caption{}
        \label{fig:phaselag_P9_coherence_original}
    \end{minipage}  
\hfill   
\setcounter{figure}{6}  
\setcounter{subfigure}{1}
    \begin{minipage}[b]{0.48\textwidth}
        \includegraphics[width=\linewidth]{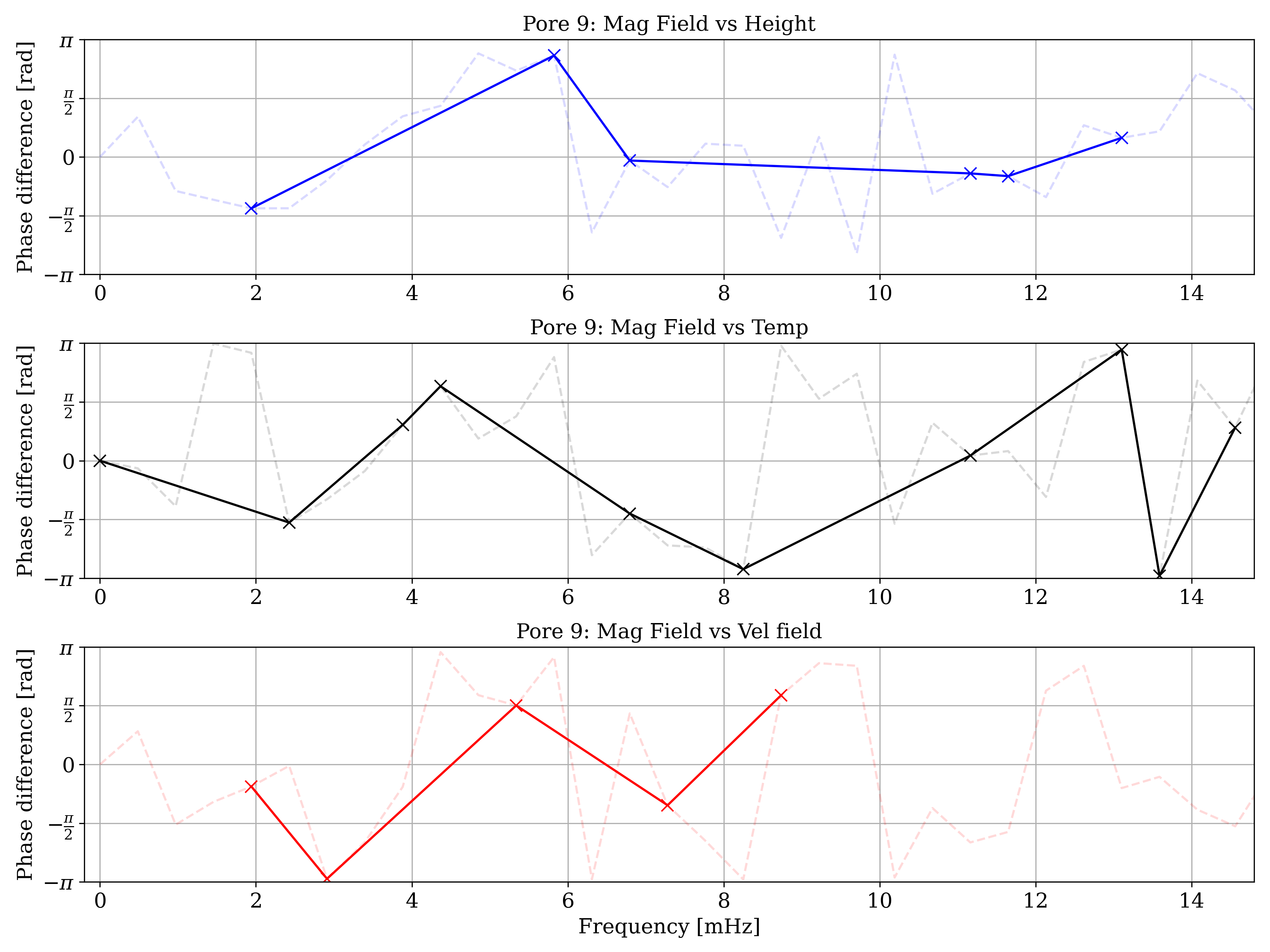}
        \caption{}
        \label{fig:phaselag_P9_coherence06}
    \end{minipage}
\setcounter{figure}{6} 
\setcounter{subfigure}{-1}
    \caption{Phase difference of \blos against (top panel) geometric height $z$, (middle panel) temperature and (bottom panel) \vlos for pore 9 at log$(\tau)=-1$. (a) Coherence threshold of 0.6. (b) Coherence threshold of 0.8. The coherent frequencies are marked with a cross and displayed with the solid line. The original signal is shown as the shaded dashed line.}
    \label{fig:phaselag_P9}
\end{subfigure}

To check if any opacity effects are tainting the signals, we conducted a phase difference analysis of the \blos against other parameters, including geometrical height and temperature using Welch's method. Figure \ref{fig:phaselag_P9} displays the phase difference between \blos and geometric height $z$, temperature and \vlos for pore 9 at the log$(\tau)=-1$ level. This plot displays the full signals, however, only the frequencies which are above a 0.8 high coherence threshold are displayed as the solid lines marked with crosses. Interestingly, at higher frequency waves ($>6$~mHz) the phase difference plateaus around $0$ degrees for the \blos vs geometrical height analysis, which suggests that these waves are not a result of opacity effects. On the other hand, the lower frequencies ($<6$~mHz) display an anti-phase/more random phase relationship, which may be a result of opacity effects.

\begin{figure}
    \centering
    \includegraphics[width=0.8\linewidth]{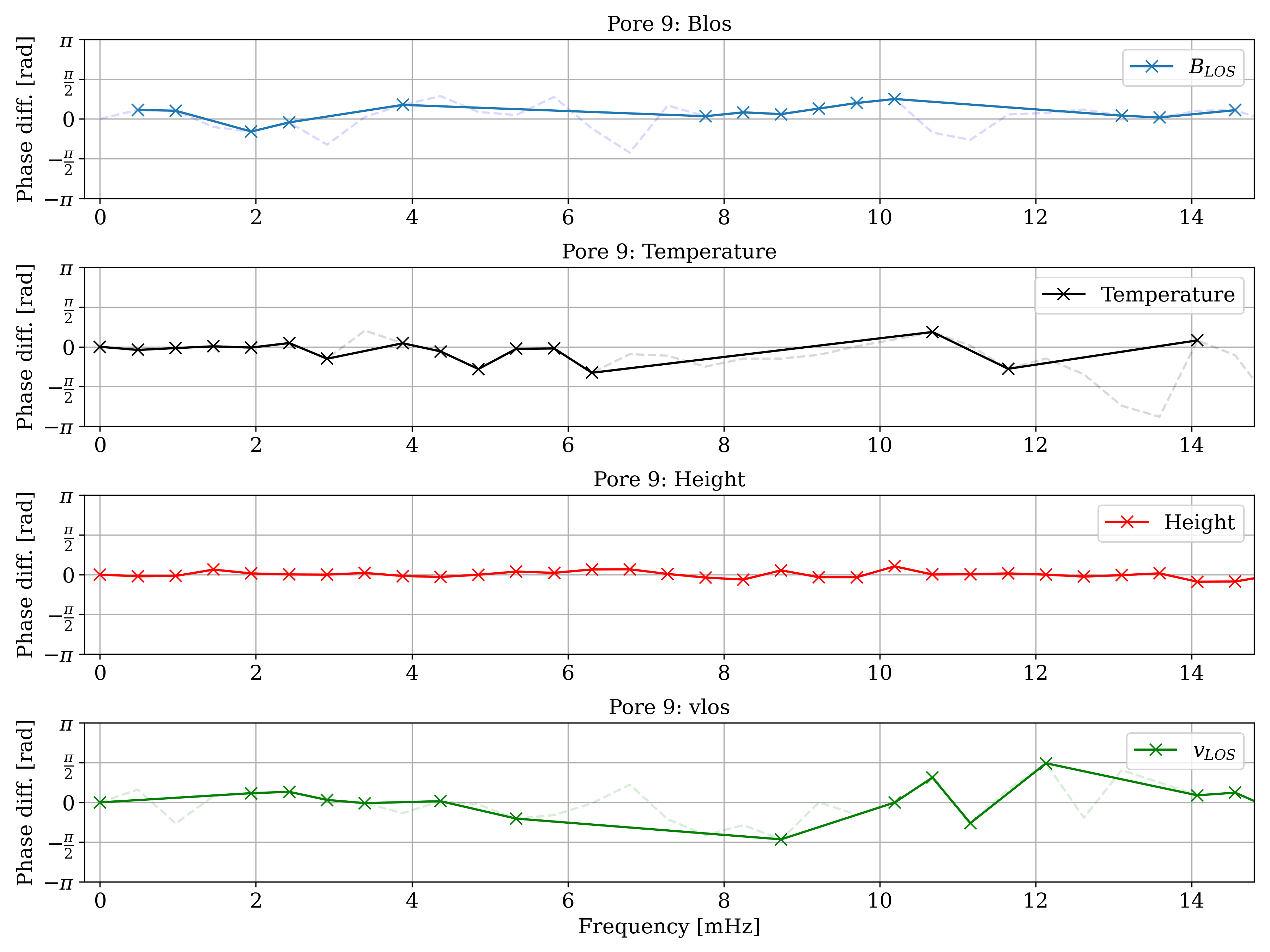}
    \caption{Phase difference between \blos, temperature, geometrical height and \vlos at log$(\tau) = -1$ and log$(\tau) = -2$ levels in Pore 9. The crosses indicate coherent frequencies in the analysis, shown with the solid line, the original signal is shown as the shaded dashed line.}
    \label{fig:pore9_phase_different_taus}
\end{figure}

Figure \ref{fig:pore9_phase_different_taus} displays the phase difference, in radians, of different parameters \blos, plasma temperature, geometrical height and \vlos in pore 9 at two different inversion levels, at log$(\tau) = -1$ and log$(\tau) = -2$. For the majority of frequencies, there is an in-phase relationship between the different levels, especially for the inferred geometrical height which displays coherent zero phase lag at all frequencies. This suggests that the height scale from the inversions is stationary and that structures within the FoV move together coherently. At higher frequencies ($>6$~mHz) there is a mainly positive phase difference in the line of sight velocity from the different log$(\tau)$ levels. Given that log$(\tau) = -2$ corresponds to a formation height higher up in the atmosphere, we can conclude that there is upward wave propagation at these higher frequencies. It could be argued that the temperature displays similar behaviour which would suggest a coupling between the \vlos and temperature, indicative of upward propagating slow magnetoacoustic waves.

It is worth noting that the anti-phase behaviour observed at low frequencies between \blos and geometric height is not uniquely attributable to opacity effects. Alternative explanations include height-dependent wave propagation delays, contributions from different magnetoacoustic wave modes, and asymmetries in the SIR inversion response functions. However, if the anti-phase \blos vs $z$ behaviour at low frequencies were due to a real propagating wave, one would expect a corresponding inter-layer phase lag in \blos between log$(\tau) = -1$ and log$(\tau) = -2$, analogous to the upward propagation signature clearly seen in \vlos at high frequencies. The absence of such a lag at low frequencies, seen in Figure \ref{fig:pore9_phase_different_taus}, where \blos, temperature and \vlos remain largely in-phase between the two inversion levels, is more consistent with a common, height-independent source such as an opacity-driven formation height shift than with genuine wave activity. Nonetheless, standing or evanescent oscillations below the acoustic cut-off would also create negligible inter-layer lags, and a conclusive determination would require dedicated forward modelling or response function analysis.

\subsection{Coherence}

To further determine whether opacity effects are present in the inversion data, we analyse the coherence and phase relationship between the signals \blos and geometric height $z$. To understand the correlation between the observed \blos oscillations and variations in $z$, given that we expect \blos to decrease with height, then opacity-induced \blos fluctuations would show a negative correlation with height. However, if instead there is a positive correlation between \blos and $z$, as reported in \citet{Joshi2018}, that would suggest that the \blos oscillations are real, and not just due to the formation height shifting (i.e., not due to opacity effect). Given the quasi-periodic nature of these signals, we adopt a wavelet coherence analysis showing when and at which frequencies the coherence is significant. The temporal information could be key to better understanding the physical origin of the observed oscillations. To conduct a wavelet analysis, a $7$x$7$ pixel square around the central pixel of the pore (i.e. $3$ pixels above, below, left and right) were used to create an average signal used for phase analysis.

\begin{subfigure}
\setcounter{figure}{8}
\setcounter{subfigure}{0}
\centering
\begin{minipage}[b]{0.47\textwidth}
    \centering
    \includegraphics[width=\linewidth]{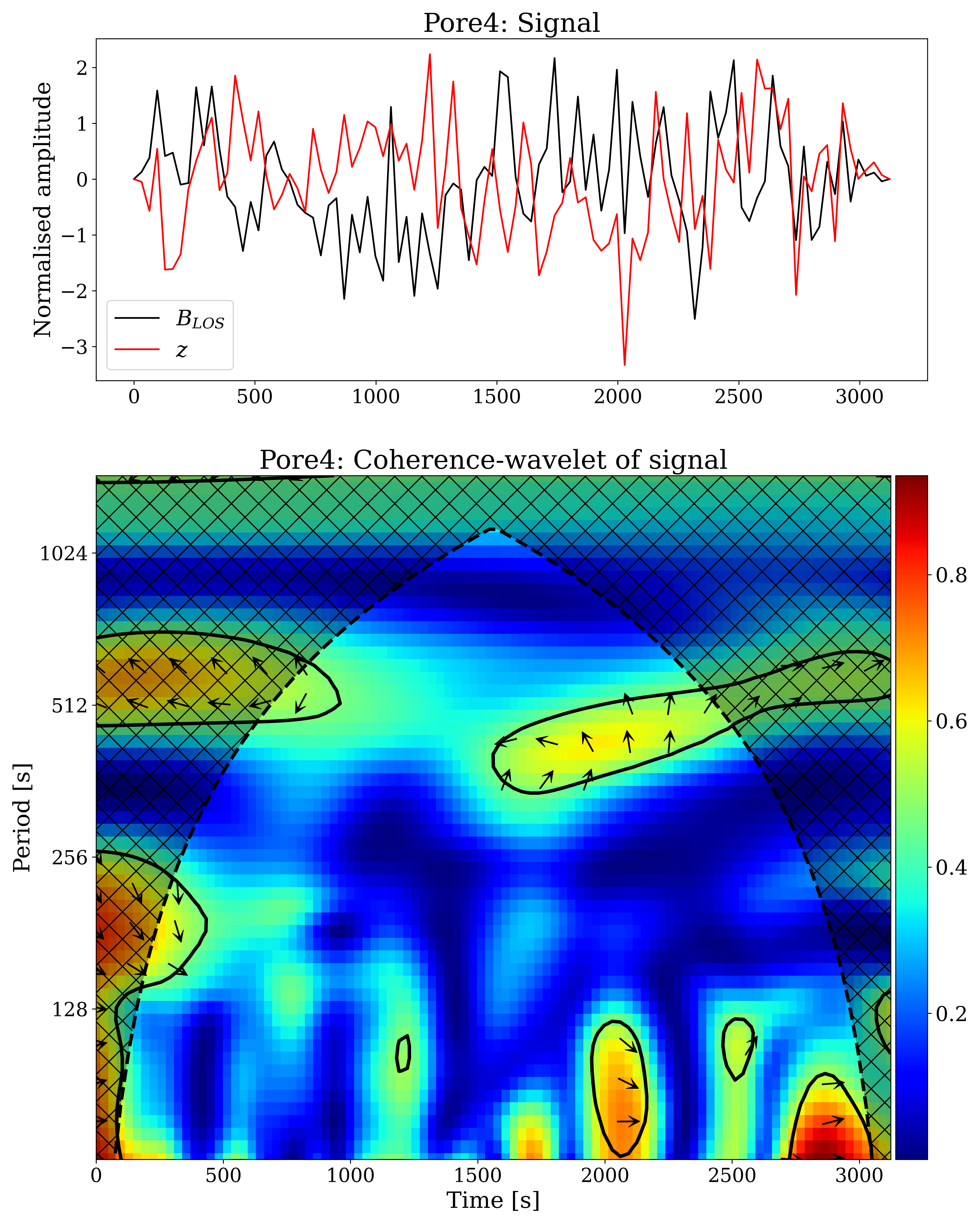}
    \caption{}
    \label{fig:wavelet_pore4}
\end{minipage}
\hfill
\setcounter{figure}{8}
\setcounter{subfigure}{1}
\begin{minipage}[b]{0.47\textwidth}
    \centering
    \includegraphics[width=\linewidth]{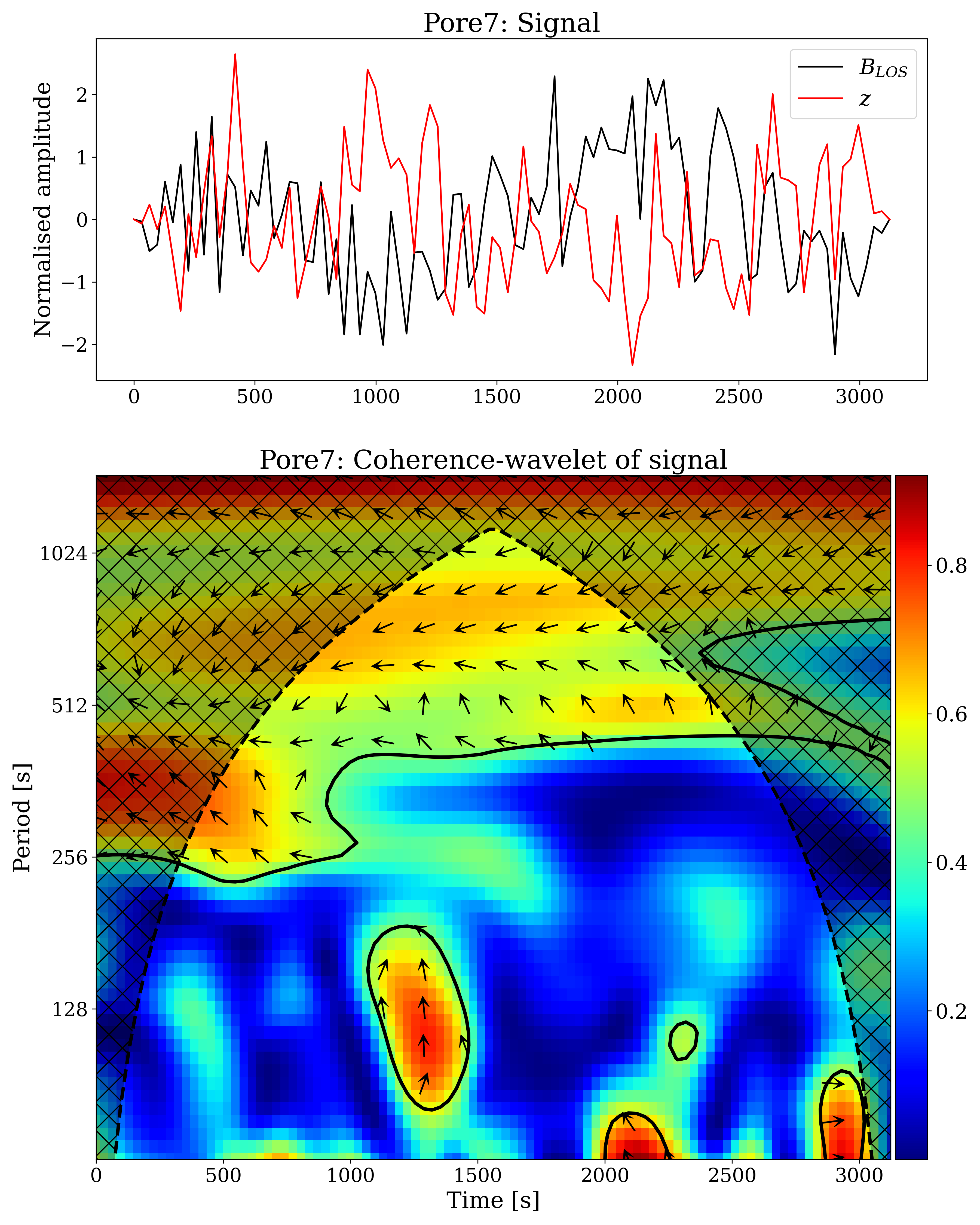}
    \caption{}
    \label{fig:wavelet_pore7}
\end{minipage}

\setcounter{figure}{8}
\setcounter{subfigure}{2}
\begin{minipage}[b]{0.47\textwidth}
    \centering
    \includegraphics[width=\linewidth]{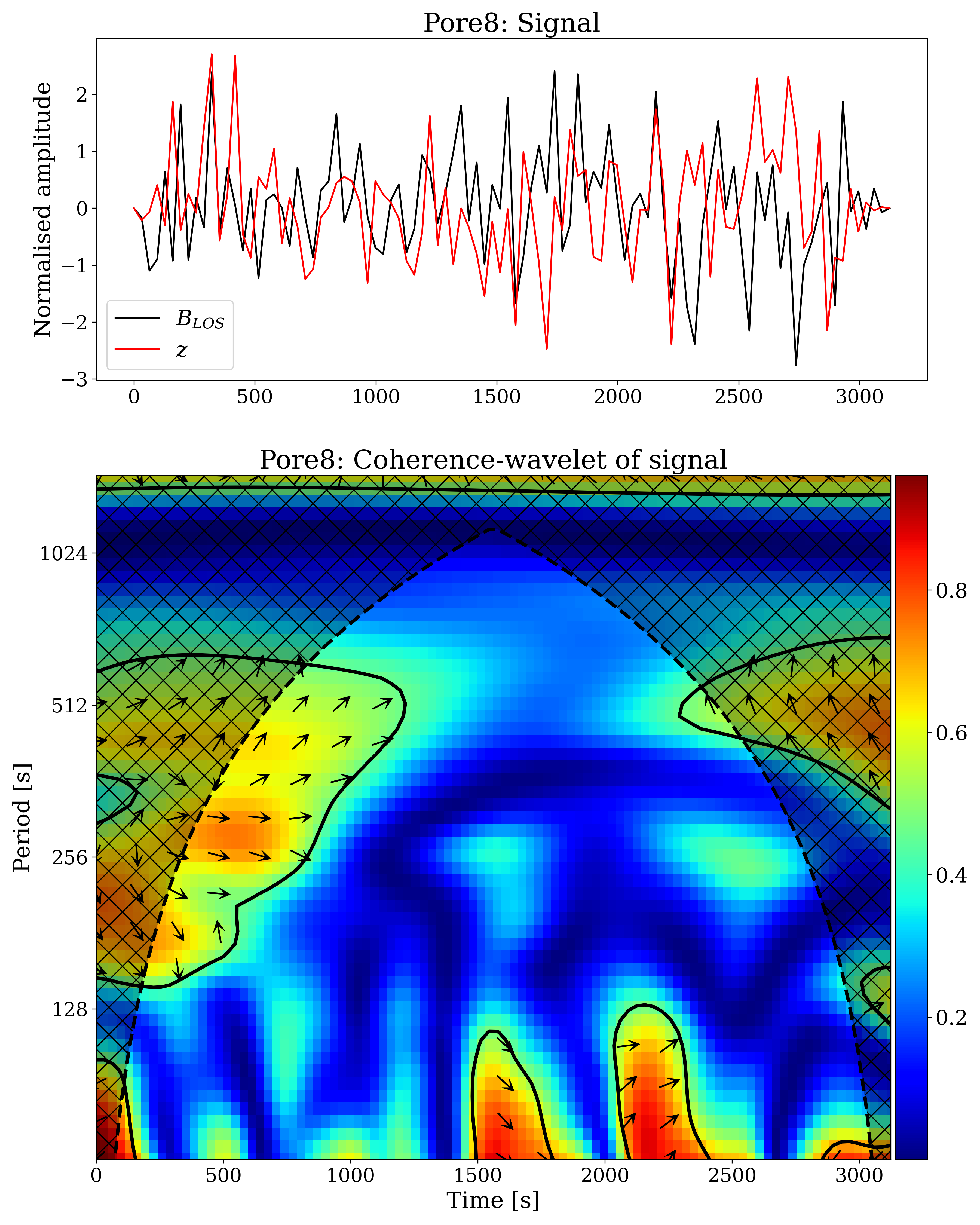}
    \caption{}
    \label{fig:wavelet_pore8}
\end{minipage}
\hfill
\setcounter{figure}{8}
\setcounter{subfigure}{3}
\begin{minipage}[b]{0.47\textwidth}
    \centering
    \includegraphics[width=\linewidth]{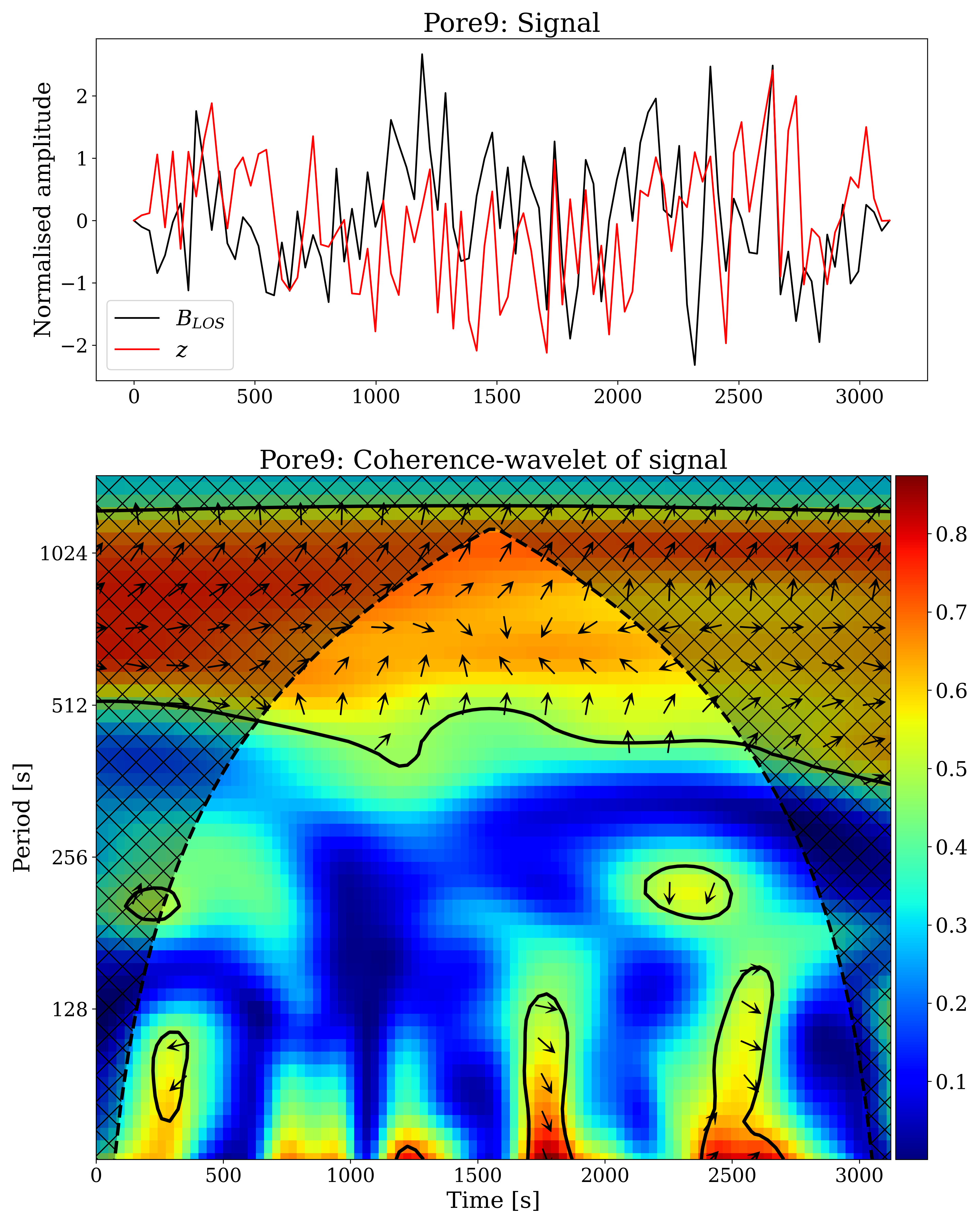}
    \caption{}
    \label{fig:wavelet_pore9}
\end{minipage}

\setcounter{figure}{8}
\setcounter{subfigure}{-1}
\caption{Wavelet coherence of $B_{\mathrm{los}}$ vs height at $\log(\tau) = -1$ for pores 4, 7, 8, and 9. Each plot shows the detrended signals and coherence-wavelet between the two signals. 
In the bottom panels, arrows indicate phase difference: $0^{\circ}$ (right), $180^{\circ}$ (left), $90^{\circ}$ (up), and $-90^{\circ}$ (down).}
\label{fig:wavelet_blos_v_z}
\end{subfigure}

Figure \ref{fig:wavelet_blos_v_z} displays the wavelet analysis between \blos and height at $\log(\tau) = -1$ for a selection of pores. It is evident that there are a range of coherent frequencies present in both signals, however there is a strong in-phase relationship at higher frequencies, i.e. frequencies above $6$~mHz (periods below $\approx 160$~s). Given the quasi-periodic nature of the signals, the in-phase relationship of higher frequencies is stronger towards the end of the observation, for times later than $1500$~s. This in phase behaviour is expected for real signatures of MHD wave propagation to the chromosphere \citep{Joshi2018}. On the other hand, the lower frequencies, typical of $p$-modes, generally display an anti-phase nature between \blos and height, which is expected if opacity effects are contaminating the signals. Therefore, we can imply that the high-frequency signals appear to be true oscillations. However, the lower ($< 3$~mHz) frequencies may be enmeshed with opacity effects, a conclusion which was also drawn by \citet{Grant2022} where, in their study, the higher frequencies seen in the chromosphere weren't coherent.

\begin{figure*}
    \centering
    \includegraphics[width=0.95\linewidth]{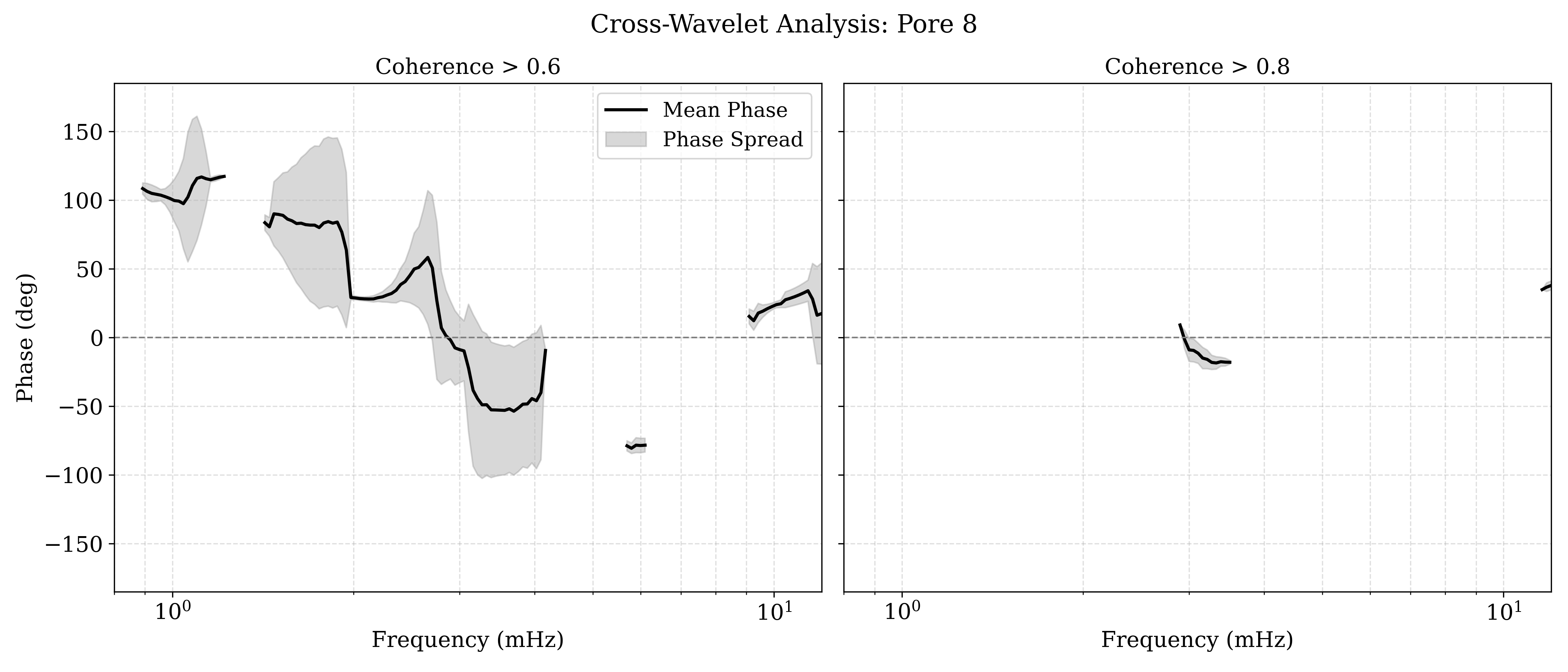}
    \caption{Coherent phase versus frequency analysis between \blos and height $z$, for pore 8 at log$(\tau) = -1$. The left panel shows the relationship between frequencies with coherence $>0.6$ whereas the right panel shows the phase vs frequencies with coherence $>0.8$. The phase spread determined using the circular standard deviation is shown as the shaded region.}
    \label{fig:pore8_logtau1_cross-wavelet}
\end{figure*}

\begin{figure*}
    \centering
    \includegraphics[width=0.95\linewidth]{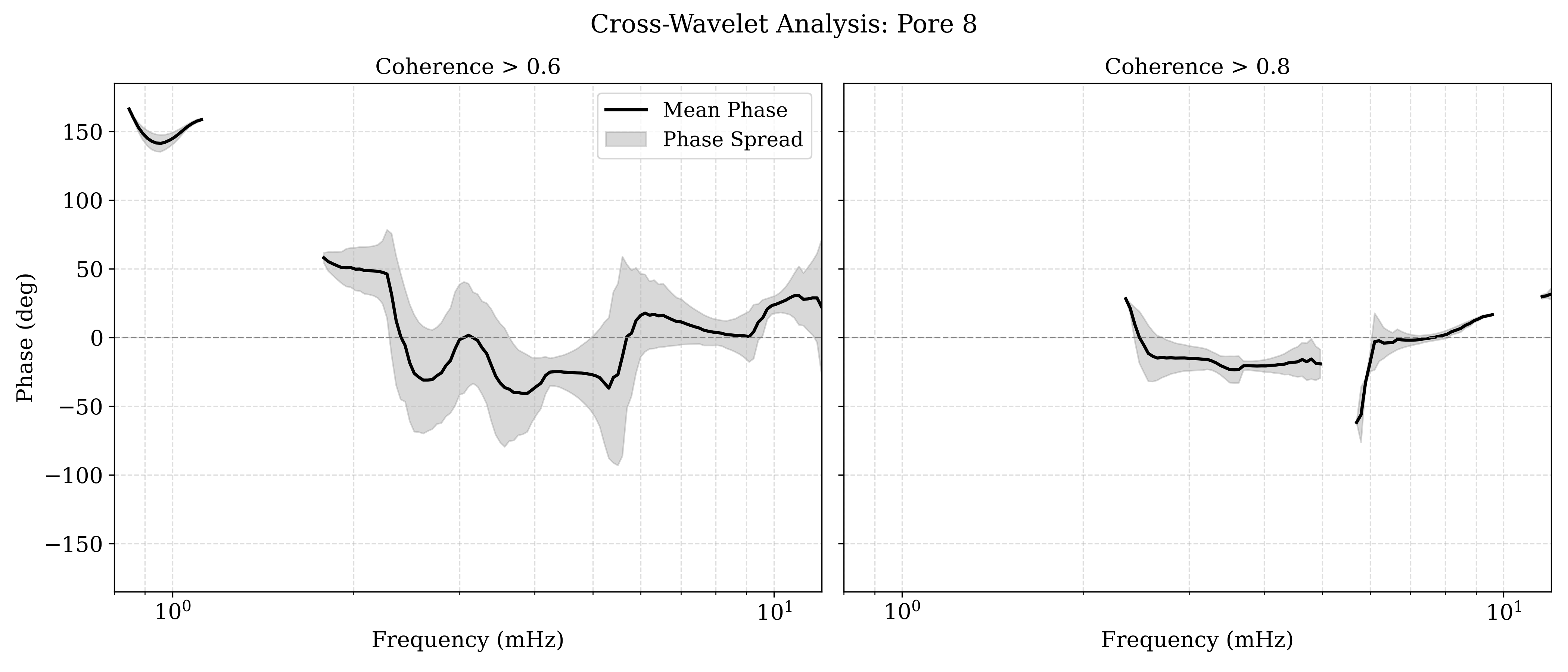}
    \caption{Same as Figure \ref{fig:pore8_logtau1_cross-wavelet} for pore 8 but for log$(\tau) = -2$.}
    \label{fig:pore8_logtau2_cross-wavelet}
\end{figure*}

To isolate the coherent frequencies, a similar cross-wavelet analysis between \blos and $z$ was conducted using WaLSAtools \citep{Jafarzadeh2025} for pore 8. Figure \ref{fig:pore8_logtau1_cross-wavelet} displays the coherent phases, for two thresholds with coherence above $0.6$ and above $0.8$ respectively, as a function of frequency for the selected pore. The shaded region in Figure \ref{fig:pore8_logtau1_cross-wavelet} highlights the circular standard deviation of the phase. This analysis further highlights the in-phase and anti-phase nature, of high ($> 6$~mHz) and low ($< 6$~mHz) frequencies, respectively, when focusing on the coherent frequencies of the two signals. We also repeated the analysis for log$(\tau) = -2$, as displayed in Figure \ref{fig:pore8_logtau2_cross-wavelet}. The same conclusion can be drawn from the behaviour of the signals at log$(\tau) = -2$ for low and high frequencies. We emphasise that the $6$~mHz division adopted here is an empirical threshold motivated by the change in phase/coherence behaviour seen in our data, rather than a universal physical cut-off. Frequencies above this value show phase signatures more consistent with upward-propagating magnetoacoustic disturbances, whereas below it the inferred \blos behaviour is more susceptible to opacity-related contamination.

\begin{figure*}
    \centering
    \includegraphics[width=0.9\linewidth]{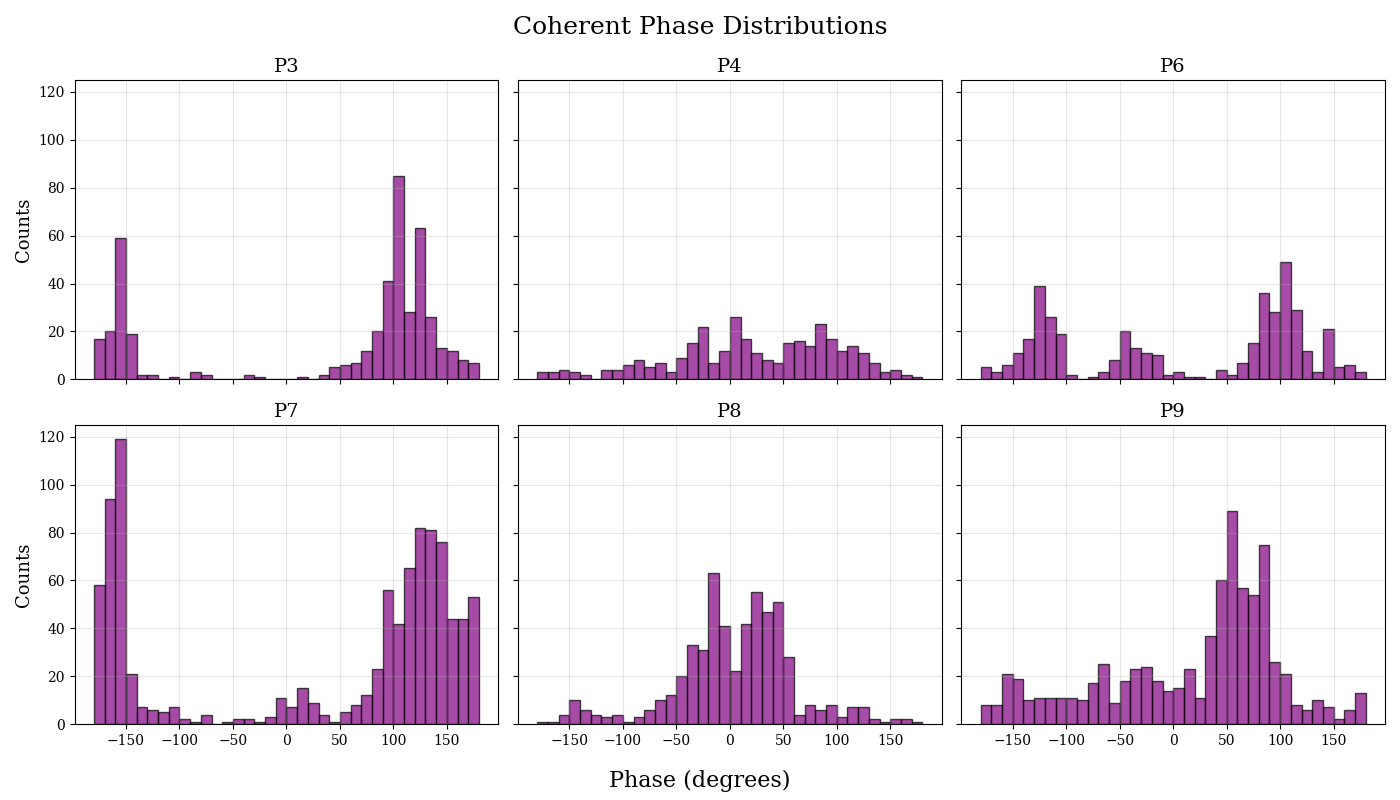}
    \caption{Histograms for the coherent phases (in degrees) between \blos and $z$ within a selection of pores from the inversions at log$(\tau) = -1$.}
    \label{fig:phase_histograms_logtau1}
\end{figure*}

\begin{figure*}
    \centering
    \includegraphics[width=0.9\linewidth]{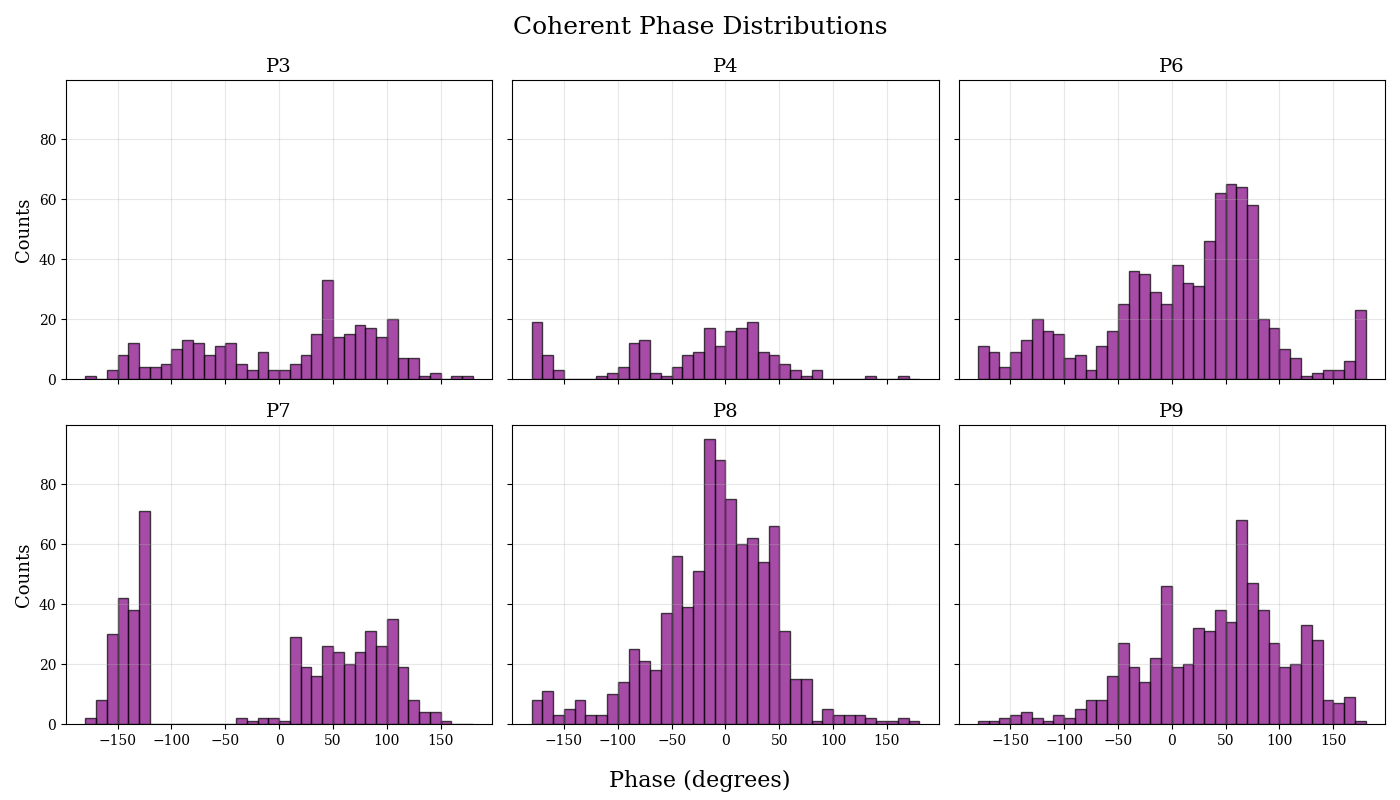}
    \caption{Same as Figure \ref{fig:phase_histograms_logtau1} but for log$(\tau) = -2$.}
    \label{fig:phase_histograms_logtau2}
\end{figure*}

Figure \ref{fig:phase_histograms_logtau1} displays histograms of coherent phase between \blos and $z$ for pores $3, 4, 6, 7, 8$ and $9$ at log$(\tau) = -1$. These histograms are essentially showing the same data denoted by the arrows outside the CoI in Figure \ref{fig:wavelet_blos_v_z} over time, reduced to a 1D array where information regarding the frequency of coherent phase is lost. The same phase distributions are shown in Figure \ref{fig:phase_histograms_logtau2} for log$(\tau) = -2$. For pores $3, 6$ and $7$ there are clear peaks towards $\pm 180^{\circ}$, suggestive of out-of-phase behaviour which indicates contamination of the signals from opacity effects. On the other hand, pore 8 displays an enhancement of phase around $0$ degrees. This behaviour in the phase analysis suggests the presence of real MHD waves. This is especially interesting as it is most pronounced in pore 8, which is also the most `rounded' pore within the SST FoV (see Figure \ref{Fig: FoV}) which indicates this pore may be the most `effective' waveguide. A similar conclusion can be drawn from the histograms of coherent phase at log$(\tau) = -2$, highlighting the coherent behaviour across the different heights spanning the peak of the Fe~{\sc{i}} $6301$~{\AA} line.

\section{Conclusions}\label{sec:conclusions}

In this study, we have examined the oscillatory behaviour within a group of solar pores using photospheric spectropolarimetric observations from CRISP at the Swedish 1-m Solar Telescope. Through SIR inversions, we retrieved the line-of-sight velocity and magnetic field, revealing strong oscillatory power at typical photospheric periods (3--5 minutes), consistent with the solar $p$-mode spectrum \citep{Lites1982}.

By analysing the SIR inversion output at multiple $\log(\tau)$ levels, we investigated how variations in \vlos, \blos, temperature, and geometrical height are interrelated. At high frequencies ($> 6$~mHz), we find coherent, in-phase fluctuations between \blos and height. Additional phase analysis at $\log(\tau) = -1$ and $\log(\tau) = -2$ indicates signatures consistent with upward-propagating magnetoacoustic waves, supported by positive phase differences in both \vlos and temperature between these layers.

A coherence-wavelet comparison of \blos and geometric height across the pore ensemble reveals clear differences: coherence and in-phase behaviour at high frequencies, but significant phase offsets at lower frequencies ($< 6$~mHz). A similar behaviour is seen across all pores over the large FoV, indicating that this may be a global phenomenon, and not isolated to an individual pore. These discrepancies highlight a critical need to verify the physical origin of low-frequency magnetic oscillations in future magnetic surveys with facilities such as {\sc Sunrise} and DKIST. In particular, the  $3$~mHz magnetic oscillations inferred from spectropolarimetric inversions may be affected by opacity and this should be further scrutinised in future studies. Ground-based facilities are particularly well-suited for these investigations, as their larger apertures enable them to resolve individual pore structures and detect the subtle polarimetric signals associated with magnetic oscillations \citep[e.g.][]{Scharmer2019, Jess2023}.

Our findings suggest that opacity effects may contaminate the observed spectropolarimetric oscillations. While the pores exhibit striking coherence in magnetic oscillations across large spatial separations, displaying coherent waveguide-like behaviour, our analysis indicates that, below $6$~mHz, opacity variations may contaminate the measured signals and artificially enhance coherence, evidenced by the greater power in \vlos and \blos for frequencies below $6$~mHz shown in Figure \ref{Pore_8_InterPore_ESD}, in addition to the change in character about $6$~mHz for the wavelet coherence analysis displayed in Figure \ref{fig:wavelet_blos_v_z}. This may partially account for the prominence of $3$~mHz power in previous studies, implying that the photospheric power distribution between $3$ and $5$~mHz may be more balanced than traditionally assumed. It is apparent that, at low frequencies below $3-4$~mHz, there is a $180^{\circ}$ phase difference between \blos and height perturbations, however, at higher frequencies, the phase between the signals tends towards $0^{\circ}$, indicative of magnetoacoustic wave phenomena.

The opacity-induced artefacts identified in this study have direct implications for numerical MHD simulations that aim to reproduce observed wave signatures in the solar atmosphere. Simulations that model wave propagation in flux tube geometries representative of waveguides in the lower solar atmosphere must account for the fact that spectropolarimetric inversions may not faithfully recover the underlying wave properties at low frequencies. In particular, simulations of slow magnetoacoustic wave propagation \citep[e.g.][]{Riedl2021, Yadav2021} and studies that forward-model Stokes diagnostics from MHD outputs \citep[e.g.][]{Felipe2023} would benefit from incorporating opacity-driven response function effects when comparing synthetic observables to real data.

To conclusively determine the extent to which opacity contributes to apparent magnetic and velocity oscillations in the lower solar atmosphere, future observational efforts must be complemented by realistic 3D numerical simulations and full radiative forward modelling. Such work will be essential to disentangle genuine MHD wave signatures from opacity-induced artefacts and to refine the interpretation of high-resolution solar magnetic diagnostics. This is vital in terms of accurately diagnosing the energy transported by MHD waves across various layers in the solar atmosphere.

\section*{Conflict of Interest Statement}

The authors declare that the research was conducted in the absence of any commercial or financial relationships that could be construed as a potential conflict of interest.

\section*{Author Contributions}

The Author Contributions section is mandatory for all articles, including articles by sole authors. If an appropriate statement is not provided on submission, a standard one will be inserted during the production process. The Author Contributions statement must describe the contributions of individual authors referred to by their initials and, in doing so, all authors agree to be accountable for the content of the work. Please see  \href{https://www.frontiersin.org/about/policies-and-publication-ethics#AuthorshipAuthorResponsibilities}{here} for full authorship criteria.

\section*{Funding}
S.J.S acknowledges support from the UKRI Future Leader Fellowship Grant (RiPSAW MR/Z000289/1).
S.J.S., S.D.T.G., D.B.J., S.J. and T.J.D. are grateful to the UK STFC for the consolidated grants ST/T00021X/1 and ST/X000923/1.
S.J. acknowledges support from the European Research Council under the European Union Horizon 2020 research and innovation program (grant agreement No. 682462) and from the Research Council of Norway through its Centres of Excellence scheme (project No. 262622). 
European Research Council under the European Union Horizon 2020 research and innovation program: grant agreement No. 682462.
Research Council of Norway Centres of Excellence scheme: project No. 262622.

\section*{Acknowledgments}

We wish to acknowledge scientific discussions with the Waves in the Lower Solar Atmosphere (WaLSA; \href{https://WaLSA.team}{www.WaLSA.team}) team, which has been supported by the Research Council of Norway (project no. 262622), The Royal Society \citep[award no. Hooke18b/SCTM;][]{2021RSPTA.37900169J}, and the International Space Science Institute (ISSI) in Bern through ISSI International Team project 502 ``WaLSA: Waves in the Lower Solar Atmosphere at High Resolution''.


\section*{Data Availability Statement}
The datasets generated for this study are available on request to the corresponding author.

\bibliographystyle{Frontiers-Harvard}
\bibliography{references}

@ARTICLE{GM21,
       author = {{Gilchrist-Millar}, Caitlin A. and {Jess}, David B. and {Grant}, Samuel D.~T. and {Keys}, Peter H.~K. and {Beck}, Christian and {Jafarzadeh}, Shahin and {Riedl}, Julia M. and {Van Doorsselaere}, Tom and {Ruiz Cobo}, Basilio},
        title = "{Magnetoacoustic wave energy dissipation in the atmosphere of solar pores}",
      journal = {Philosophical Transactions of the Royal Society of London Series A},
         year = 2021,
        month = feb,
       volume = {379},
       number = {2190},
          eid = {20200172},
        pages = {20200172},
          doi = {10.1098/rsta.2020.0172},
archivePrefix = {arXiv},
       eprint = {2007.11594},
 primaryClass = {astro-ph.SR},
       adsurl = {https://ui.adsabs.harvard.edu/abs/2021RSPTA.37900172G}
}

@ARTICLE{2021RSPTA.37900169J,
       author = {{Jess}, D.~B. and {Keys}, P.~H. and {Stangalini}, M. and {Jafarzadeh}, S.},
        title = "{High-resolution wave dynamics in the lower solar atmosphere}",
      journal = {Philosophical Transactions of the Royal Society of London Series A},
         year = 2021,
        month = feb,
       volume = {379},
       number = {2190},
          eid = {20200169},
        pages = {20200169},
          doi = {10.1098/rsta.2020.0169},
archivePrefix = {arXiv},
       eprint = {2011.13940},
 primaryClass = {astro-ph.SR},
       adsurl = {https://ui.adsabs.harvard.edu/abs/2021RSPTA.37900169J}
}

@ARTICLE{Nelson2021,
       author = {{Nelson}, C.~J. and {Campbell}, R.~J. and {Mathioudakis}, M.},
        title = "{Oscillations in the line-of-sight magnetic field strength in a pore observed by the GREGOR Infrared Spectrograph (GRIS)}",
      journal = {Astronomy \& Astrophysics},
         year = 2021,
        month = oct,
       volume = {654},
          eid = {A50},
        pages = {A50},
          doi = {10.1051/0004-6361/202141368},
archivePrefix = {arXiv},
       eprint = {2107.10183},
 primaryClass = {astro-ph.SR},
       adsurl = {https://ui.adsabs.harvard.edu/abs/2021A&A...654A..50N}
}

@article{SUI20122159,
title = {Four Methods for Roundness Evaluation},
journal = {Physics Procedia},
volume = {24},
pages = {2159-2164},
year = {2012},
note = {International Conference on Applied Physics and Industrial Engineering 2012},
issn = {1875-3892},
doi = {https://doi.org/10.1016/j.phpro.2012.02.317},
url = {https://www.sciencedirect.com/science/article/pii/S1875389212003604},
author = {Wentao Sui and Dan Zhang}}

@ARTICLE{Felipe2023,
       author = {{Felipe}, T. and {Socas-Navarro}, H.},
        title = "{Impact of opacity effects on chromospheric oscillations inferred from NLTE inversions}",
      journal = {Astronomy \& Astrophysics},
         year = 2023,
        month = feb,
       volume = {670},
          eid = {A133},
        pages = {A133},
          doi = {10.1051/0004-6361/202245439},
archivePrefix = {arXiv},
       eprint = {2301.03273},
 primaryClass = {astro-ph.SR},
       adsurl = {https://ui.adsabs.harvard.edu/abs/2023A&A...670A.133F}
}

@ARTICLE{Joshi2018,
       author = {{Joshi}, Jayant and {de la Cruz Rodr{\'\i}guez}, Jaime},
        title = "{Magnetic field variations associated with umbral flashes and penumbral waves}",
      journal = {Astronomy \& Astrophysics},
         year = 2018,
        month = nov,
       volume = {619},
          eid = {A63},
        pages = {A63},
          doi = {10.1051/0004-6361/201832955},
archivePrefix = {arXiv},
       eprint = {1803.01737},
 primaryClass = {astro-ph.SR},
       adsurl = {https://ui.adsabs.harvard.edu/abs/2018A&A...619A..63J}
}

@ARTICLE{Stan2021,
       author = {{Stangalini}, M. and {Baker}, D. and {Valori}, G. and {Jess}, D.~B. and {Jafarzadeh}, S. and {Murabito}, M. and {To}, A.~S.~H. and {Brooks}, D.~H. and {Ermolli}, I. and {Giorgi}, F. and {MacBride}, C.~D.},
        title = "{Spectropolarimetric fluctuations in a sunspot chromosphere}",
      journal = {Philosophical Transactions of the Royal Society of London Series A},
         year = 2021,
        month = feb,
       volume = {379},
       number = {2190},
          eid = {20200216},
        pages = {20200216},
          doi = {10.1098/rsta.2020.0216},
archivePrefix = {arXiv},
       eprint = {2009.05302},
 primaryClass = {astro-ph.SR},
       adsurl = {https://ui.adsabs.harvard.edu/abs/2021RSPTA.37900216S}
}

@ARTICLE{Fujimura2009,
       author = {{Fujimura}, D. and {Tsuneta}, S.},
        title = "{Properties of Magnetohydrodynamic Waves in the Solar Photosphere Obtained with Hinode}",
      journal = {The Astrophysical Journal},
         year = 2009,
        month = sep,
       volume = {702},
       number = {2},
        pages = {1443-1457},
          doi = {10.1088/0004-637X/702/2/1443},
archivePrefix = {arXiv},
       eprint = {0907.3025},
 primaryClass = {astro-ph.SR},
       adsurl = {https://ui.adsabs.harvard.edu/abs/2009ApJ...702.1443F}
}

@ARTICLE{Jaf2024,
       author = {{Jafarzadeh}, S. and {Schiavo}, L.~A.~C.~A. and {Fedun}, V. and {Solanki}, S.~K. and {Stangalini}, M. and {Calchetti}, D. and {Verth}, G. and {Jess}, D.~B. and {Grant}, S.~D.~T. and {Ballai}, I. and {Gafeira}, R. and {Keys}, P.~H. and {Fleck}, B. and {Morton}, R.~J. and {Browning}, P.~K. and {Silva}, S.~S.~A. and {Appourchaux}, T. and {Gandorfer}, A. and {Gizon}, L. and {Hirzberger}, J. and {Kahil}, F. and {Orozco Su{\'a}rez}, D. and {Schou}, J. and {Strecker}, H. and {del Toro Iniesta}, J.~C. and {Valori}, G. and {Volkmer}, R. and {Woch}, J.},
        title = "{Sausage, kink, and fluting magnetohydrodynamic wave modes identified in solar magnetic pores by Solar Orbiter/PHI}",
      journal = {Astronomy \& Astrophysics},
         year = 2024,
        month = aug,
       volume = {688},
          eid = {A2},
        pages = {A2},
          doi = {10.1051/0004-6361/202449685},
archivePrefix = {arXiv},
       eprint = {2404.18717},
 primaryClass = {astro-ph.SR},
       adsurl = {https://ui.adsabs.harvard.edu/abs/2024A&A...688A...2J}
}

@ARTICLE{Murabito2021,
       author = {{Murabito}, M. and {Stangalini}, M. and {Baker}, D. and {Valori}, G. and {Jess}, D.~B. and {Jafarzadeh}, S. and {Brooks}, D.~H. and {Ermolli}, I. and {Giorgi}, F. and {Grant}, S.~D.~T. and {Long}, D.~M. and {van Driel-Gesztelyi}, L.},
        title = "{Investigating the origin of magnetic perturbations associated with the FIP Effect}",
      journal = {Astronomy \& Astrophysics},
         year = 2021,
        month = dec,
       volume = {656},
          eid = {A87},
        pages = {A87},
          doi = {10.1051/0004-6361/202141504},
archivePrefix = {arXiv},
       eprint = {2108.11164},
 primaryClass = {astro-ph.SR},
       adsurl = {https://ui.adsabs.harvard.edu/abs/2021A&A...656A..87M}
}

@ARTICLE{Houston2018,
       author = {{Houston}, S.~J. and {Jess}, D.~B. and {Asensio Ramos}, A. and {Grant}, S.~D.~T. and {Beck}, C. and {Norton}, A.~A. and {Krishna Prasad}, S.},
        title = "{The Magnetic Response of the Solar Atmosphere to Umbral Flashes}",
      journal = {The Astrophysical Journal},
         year = 2018,
        month = jun,
       volume = {860},
       number = {1},
          eid = {28},
        pages = {28},
          doi = {10.3847/1538-4357/aab366},
archivePrefix = {arXiv},
       eprint = {1803.00018},
 primaryClass = {astro-ph.SR},
       adsurl = {https://ui.adsabs.harvard.edu/abs/2018ApJ...860...28H}
}

@ARTICLE{Bailen2023,
       author = {{Bail{\'e}n}, F.~J. and {Orozco Su{\'a}rez}, D. and {del Toro Iniesta}, J.~C.},
        title = "{Fabry-P{\'e}rot etalons in solar astronomy. A review}",
      journal = {Astrophysics and Space Science},
         year = 2023,
        month = jul,
       volume = {368},
       number = {7},
          eid = {55},
        pages = {55},
          doi = {10.1007/s10509-023-04212-3},
       adsurl = {https://ui.adsabs.harvard.edu/abs/2023Ap&SS.368...55B}
}

@ARTICLE{Rod2017,
       author = {{de la Cruz Rodr{\'\i}guez}, J. and {van Noort}, M.},
        title = "{Radiative Diagnostics in the Solar Photosphere and Chromosphere}",
      journal = {Space Science Reviews},
         year = 2017,
        month = sep,
       volume = {210},
       number = {1-4},
        pages = {109-143},
          doi = {10.1007/s11214-016-0294-8},
archivePrefix = {arXiv},
       eprint = {1609.08324},
 primaryClass = {astro-ph.SR},
       adsurl = {https://ui.adsabs.harvard.edu/abs/2017SSRv..210..109D}
}

@ARTICLE{Stangalini2021,
       author = {{Stangalini}, M. and {Jess}, D.~B. and {Verth}, G. and {Fedun}, V. and {Fleck}, B. and {Jafarzadeh}, S. and {Keys}, P.~H. and {Murabito}, M. and {Calchetti}, D. and {Aldhafeeri}, A.~A. and {Berrilli}, F. and {Del Moro}, D. and {Jefferies}, S.~M. and {Terradas}, J. and {Soler}, R.},
        title = "{A novel approach to identify resonant MHD wave modes in solar pores and sunspot umbrae: B {\ensuremath{-}} {\ensuremath{\omega}} analysis}",
      journal = {Astronomy \& Astrophysics},
         year = 2021,
        month = may,
       volume = {649},
          eid = {A169},
        pages = {A169},
          doi = {10.1051/0004-6361/202140429},
archivePrefix = {arXiv},
       eprint = {2103.11639},
 primaryClass = {astro-ph.SR},
       adsurl = {https://ui.adsabs.harvard.edu/abs/2021A&A...649A.169S}
}

@ARTICLE{Morton2011,
       author = {{Morton}, R.~J. and {Erd{\'e}lyi}, R. and {Jess}, D.~B. and {Mathioudakis}, M.},
        title = "{Observations of Sausage Modes in Magnetic Pores}",
      journal = {The Astrophysical Journal Letters},
         year = 2011,
        month = mar,
       volume = {729},
       number = {2},
          eid = {L18},
        pages = {L18},
          doi = {10.1088/2041-8205/729/2/L18},
archivePrefix = {arXiv},
       eprint = {1011.2375},
 primaryClass = {astro-ph.SR},
       adsurl = {https://ui.adsabs.harvard.edu/abs/2011ApJ...729L..18M}
}

@ARTICLE{Keys2018,
       author = {{Keys}, Peter H. and {Morton}, Richard J. and {Jess}, David B. and {Verth}, Gary and {Grant}, Samuel D.~T. and {Mathioudakis}, Mihalis and {Mackay}, Duncan H. and {Doyle}, John G. and {Christian}, Damian J. and {Keenan}, Francis P. and {Erd{\'e}lyi}, Robertus},
        title = "{Photospheric Observations of Surface and Body Modes in Solar Magnetic Pores}",
      journal = {The Astrophysical Journal},
         year = 2018,
        month = apr,
       volume = {857},
       number = {1},
          eid = {28},
        pages = {28},
          doi = {10.3847/1538-4357/aab432},
archivePrefix = {arXiv},
       eprint = {1803.01859},
 primaryClass = {astro-ph.SR},
       adsurl = {https://ui.adsabs.harvard.edu/abs/2018ApJ...857...28K}
}

@ARTICLE{Grant2022,
       author = {{Grant}, S.~D.~T. and {Jess}, D.~B. and {Stangalini}, M. and {Jafarzadeh}, S. and {Fedun}, V. and {Verth}, G. and {Keys}, P.~H. and {Rajaguru}, S.~P. and {Uitenbroek}, H. and {MacBride}, C.~D. and {Bate}, W. and {Gilchrist-Millar}, C.~A.},
        title = "{The Propagation of Coherent Waves Across Multiple Solar Magnetic Pores}",
      journal = {The Astrophysical Journal},
         year = 2022,
        month = oct,
       volume = {938},
       number = {2},
          eid = {143},
        pages = {143},
          doi = {10.3847/1538-4357/ac91ca},
archivePrefix = {arXiv},
       eprint = {2209.06280},
 primaryClass = {astro-ph.SR},
       adsurl = {https://ui.adsabs.harvard.edu/abs/2022ApJ...938..143G}
}

@ARTICLE{VD2020,
       author = {{Van Doorsselaere}, Tom and {Srivastava}, Abhishek K. and {Antolin}, Patrick and {Magyar}, Norbert and {Vasheghani Farahani}, Soheil and {Tian}, Hui and {Kolotkov}, Dmitrii and {Ofman}, Leon and {Guo}, Mingzhe and {Arregui}, I{\~n}igo and {De Moortel}, Ineke and {Pascoe}, David},
        title = "{Coronal Heating by MHD Waves}",
      journal = {Space Science Reviews},
         year = 2020,
        month = dec,
       volume = {216},
       number = {8},
          eid = {140},
        pages = {140},
          doi = {10.1007/s11214-020-00770-y},
archivePrefix = {arXiv},
       eprint = {2012.01371},
 primaryClass = {astro-ph.SR},
       adsurl = {https://ui.adsabs.harvard.edu/abs/2020SSRv..216..140V}
}

@ARTICLE{Jess2023,
       author = {{Jess}, David B. and {Jafarzadeh}, Shahin and {Keys}, Peter H. and {Stangalini}, Marco and {Verth}, Gary and {Grant}, Samuel D.~T.},
        title = "{Waves in the lower solar atmosphere: the dawn of next-generation solar telescopes}",
      journal = {Living Reviews in Solar Physics},
         year = 2023,
        month = dec,
       volume = {20},
       number = {1},
          eid = {1},
        pages = {1},
          doi = {10.1007/s41116-022-00035-6},
archivePrefix = {arXiv},
       eprint = {2212.09788},
 primaryClass = {astro-ph.SR},
       adsurl = {https://ui.adsabs.harvard.edu/abs/2023LRSP...20....1J}
}

@ARTICLE{Jess2015,
       author = {{Jess}, D.~B. and {Morton}, R.~J. and {Verth}, G. and {Fedun}, V. and {Grant}, S.~D.~T. and {Giagkiozis}, I.},
        title = "{Multiwavelength Studies of MHD Waves in the Solar Chromosphere. An Overview of Recent Results}",
      journal = {Space Science Reviews},
         year = 2015,
        month = jul,
       volume = {190},
       number = {1-4},
        pages = {103-161},
          doi = {10.1007/s11214-015-0141-3},
archivePrefix = {arXiv},
       eprint = {1503.01769},
 primaryClass = {astro-ph.SR},
       adsurl = {https://ui.adsabs.harvard.edu/abs/2015SSRv..190..103J}
}

@INPROCEEDINGS{Doro2008,
       author = {{Dorotovi{\v{c}}}, I. and {Erd{\'e}lyi}, R. and {Karlovsk{\'y}}, V.},
        title = "{Identification of linear slow sausage waves in magnetic pores}",
    booktitle = {Waves \& Oscillations in the Solar Atmosphere: Heating and Magneto-Seismology},
         year = 2008,
       editor = {{Erd{\'e}lyi}, Robert and {Mendoza-Briceno}, C{\'e}sar A.},
       volume = {247},
        month = may,
        pages = {351-354},
          doi = {10.1017/S174392130801507X},
       adsurl = {https://ui.adsabs.harvard.edu/abs/2008IAUS..247..351D}
}

@ARTICLE{Grant2015,
       author = {{Grant}, S.~D.~T. and {Jess}, D.~B. and {Moreels}, M.~G. and {Morton}, R.~J. and {Christian}, D.~J. and {Giagkiozis}, I. and {Verth}, G. and {Fedun}, V. and {Keys}, P.~H. and {Van Doorsselaere}, T. and {Erd{\'e}lyi}, R.},
        title = "{Wave Damping Observed in Upwardly Propagating Sausage-mode Oscillations Contained within a Magnetic Pore}",
      journal = {The Astrophysical Journal},
         year = 2015,
        month = jun,
       volume = {806},
       number = {1},
          eid = {132},
        pages = {132},
          doi = {10.1088/0004-637X/806/1/132},
archivePrefix = {arXiv},
       eprint = {1505.01484},
 primaryClass = {astro-ph.SR},
       adsurl = {https://ui.adsabs.harvard.edu/abs/2015ApJ...806..132G}
}

@ARTICLE{Riedl2021,
       author = {{Riedl}, J.~M. and {Gilchrist-Millar}, C.~A. and {Van Doorsselaere}, T. and {Jess}, D.~B. and {Grant}, S.~D.~T.},
        title = "{Finding the mechanism of wave energy flux damping in solar pores using numerical simulations}",
      journal = {Astronomy \& Astrophysics},
         year = 2021,
        month = apr,
       volume = {648},
          eid = {A77},
        pages = {A77},
          doi = {10.1051/0004-6361/202040163},
archivePrefix = {arXiv},
       eprint = {2102.12420},
 primaryClass = {astro-ph.SR},
       adsurl = {https://ui.adsabs.harvard.edu/abs/2021A&A...648A..77R}
}

@INPROCEEDINGS{Scharmer2003,
       author = {{Scharmer}, Goran B. and {Bjelksjo}, Klas and {Korhonen}, Tapio K. and {Lindberg}, Bo and {Petterson}, Bertil},
        title = "{The 1-meter Swedish solar telescope}",
    booktitle = {Innovative Telescopes and Instrumentation for Solar Astrophysics},
         year = 2003,
       editor = {{Keil}, Stephen L. and {Avakyan}, Sergey V.},
       series = {Society of Photo-Optical Instrumentation Engineers (SPIE) Conference Series},
       volume = {4853},
        month = feb,
        pages = {341-350},
          doi = {10.1117/12.460377},
       adsurl = {https://ui.adsabs.harvard.edu/abs/2003SPIE.4853..341S}
}

@ARTICLE{Scharmer2019,
       author = {{Scharmer}, G.~B. and {L{\"o}fdahl}, M.~G. and {Sliepen}, G. and {de la Cruz Rodr{\'\i}guez}, J.},
        title = "{Is the sky the limit?. Performance of the revamped Swedish 1-m Solar Telescope and its blue- and red-beam reimaging systems}",
      journal = {Astronomy \& Astrophysics},
         year = 2019,
        month = jun,
       volume = {626},
          eid = {A55},
        pages = {A55},
          doi = {10.1051/0004-6361/201935735},
archivePrefix = {arXiv},
       eprint = {1905.05588},
 primaryClass = {astro-ph.IM},
       adsurl = {https://ui.adsabs.harvard.edu/abs/2019A&A...626A..55S}
}

@ARTICLE{Rodriguez2015,
       author = {{de la Cruz Rodr{\'\i}guez}, J. and {L{\"o}fdahl}, M.~G. and {S{\"u}tterlin}, P. and {Hillberg}, T. and {Rouppe van der Voort}, L.},
        title = "{CRISPRED: A data pipeline for the CRISP imaging spectropolarimeter}",
      journal = {Astronomy \& Astrophysics},
         year = 2015,
        month = jan,
       volume = {573},
          eid = {A40},
        pages = {A40},
          doi = {10.1051/0004-6361/201424319},
archivePrefix = {arXiv},
       eprint = {1406.0202},
 primaryClass = {astro-ph.SR},
       adsurl = {https://ui.adsabs.harvard.edu/abs/2015A&A...573A..40D}
}

@ARTICLE{Scharmer2008,
       author = {{Scharmer}, G.~B. and {Narayan}, G. and {Hillberg}, T. and {de la Cruz Rodriguez}, J. and {L{\"o}fdahl}, M.~G. and {Kiselman}, D. and {S{\"u}tterlin}, P. and {van Noort}, M. and {Lagg}, A.},
        title = "{CRISP Spectropolarimetric Imaging of Penumbral Fine Structure}",
      journal = {The Astrophysical Journal Letters},
         year = 2008,
        month = dec,
       volume = {689},
       number = {1},
        pages = {L69},
          doi = {10.1086/595744},
archivePrefix = {arXiv},
       eprint = {0806.1638},
 primaryClass = {astro-ph},
       adsurl = {https://ui.adsabs.harvard.edu/abs/2008ApJ...689L..69S}
}

@INPROCEEDINGS{Scharmer2003b,
       author = {{Scharmer}, Goran B. and {Dettori}, Peter M. and {Lofdahl}, Mats G. and {Shand}, Mark},
        title = "{Adaptive optics system for the new Swedish solar telescope}",
    booktitle = {Innovative Telescopes and Instrumentation for Solar Astrophysics},
         year = 2003,
       editor = {{Keil}, Stephen L. and {Avakyan}, Sergey V.},
       series = {Society of Photo-Optical Instrumentation Engineers (SPIE) Conference Series},
       volume = {4853},
        month = feb,
        pages = {370-380},
          doi = {10.1117/12.460387},
       adsurl = {https://ui.adsabs.harvard.edu/abs/2003SPIE.4853..370S}
}

@ARTICLE{Noort2005,
       author = {{Van Noort}, Michiel and {Rouppe Van Der Voort}, Luc and {L{\"o}fdahl}, Mats G.},
        title = "{Solar Image Restoration By Use Of Multi-frame Blind De-convolution With Multiple Objects And Phase Diversity}",
      journal = {Solar Physics},
         year = 2005,
        month = may,
       volume = {228},
       number = {1-2},
        pages = {191-215},
          doi = {10.1007/s11207-005-5782-z},
       adsurl = {https://ui.adsabs.harvard.edu/abs/2005SoPh..228..191V}
}

@ARTICLE{Rozo2023,
       author = {{Campos Rozo}, J.~I. and {Vargas Dom{\'\i}nguez}, S. and {Utz}, D. and {Veronig}, A.~M. and {Hanslmeier}, A.},
        title = "{Exploring magnetic field properties at the boundary of solar pores: A comparative study based on SDO-HMI observations}",
      journal = {Astronomy \& Astrophysics},
         year = 2023,
        month = jun,
       volume = {674},
          eid = {A91},
        pages = {A91},
          doi = {10.1051/0004-6361/202346389},
archivePrefix = {arXiv},
       eprint = {2304.13212},
 primaryClass = {astro-ph.SR},
       adsurl = {https://ui.adsabs.harvard.edu/abs/2023A&A...674A..91C}
}

@ARTICLE{Garcia1987,
       author = {{Garcia de La Rosa}, J.~I.},
        title = "{Umbral Dots - a Case of Penetrative Convection Between Sunspot Fragments}",
      journal = {Solar Physics},
         year = 1987,
        month = mar,
       volume = {112},
       number = {1},
        pages = {49-58},
          doi = {10.1007/BF00148486},
       adsurl = {https://ui.adsabs.harvard.edu/abs/1987SoPh..112...49G}
}

@ARTICLE{Sob2003,
       author = {{Sobotka}, M.},
        title = "{Solar activity II: Sunspots and pores}",
      journal = {Astronomische Nachrichten},
         year = 2003,
        month = jan,
       volume = {324},
       number = {4},
        pages = {369-373},
          doi = {10.1002/asna.200310132},
       adsurl = {https://ui.adsabs.harvard.edu/abs/2003AN....324..369S}
}

@ARTICLE{Torrence1998,
       author = {{Torrence}, Christopher and {Compo}, Gilbert P.},
        title = "{A Practical Guide to Wavelet Analysis.}",
      journal = {Bulletin of the American Meteorological Society},
         year = 1998,
        month = jan,
       volume = {79},
       number = {1},
        pages = {61-78},
          doi = {10.1175/1520-0477(1998)079<0061:APGTWA>2.0.CO;2},
       adsurl = {https://ui.adsabs.harvard.edu/abs/1998BAMS...79...61T}
}

@ARTICLE{Lites1982,
       author = {{Lites}, B.~W. and {White}, O.~R. and {Packman}, D.},
        title = "{Photoelectric observations of propagating sunspot oscillations}",
      journal = {The Astrophysical Journal},
         year = 1982,
        month = feb,
       volume = {253},
        pages = {386-392},
          doi = {10.1086/159642},
       adsurl = {https://ui.adsabs.harvard.edu/abs/1982ApJ...253..386L}
}

@ARTICLE{Ruiz1992,
       author = {{Ruiz Cobo}, B. and {del Toro Iniesta}, J.~C.},
        title = "{Inversion of Stokes Profiles}",
      journal = {The Astrophysical Journal},
         year = 1992,
        month = oct,
       volume = {398},
        pages = {375},
          doi = {10.1086/171862},
       adsurl = {https://ui.adsabs.harvard.edu/abs/1992ApJ...398..375R}
}

@ARTICLE{Jafarzadeh2025,
       author = {{Jafarzadeh}, Shahin and {Jess}, David B. and {Stangalini}, Marco and {Grant}, Samuel D.~T. and {Higham}, Jonathan E. and {Pessah}, Martin E. and {Keys}, Peter H. and {Belov}, Sergey and {Calchetti}, Daniele and {Duckenfield}, Timothy J. and {Fedun}, Viktor and {Fleck}, Bernhard and {Gafeira}, Ricardo and {Jefferies}, Stuart M. and {Khomenko}, Elena and {Morton}, Richard J. and {Norton}, Aimee A. and {Rajaguru}, S.~P. and {Schiavo}, Luiz A.~C.~A. and {Sharma}, Rahul and {Silva}, Suzana S.~A. and {Solanki}, Sami K. and {Steiner}, Oskar and {Verth}, Gary and {Vigeesh}, Gangadharan and {Yadav}, Nitin},
        title = "{Wave analysis tools}",
      journal = {Nature Reviews Methods Primers},
         year = 2025,
        month = apr,
       volume = {5},
          eid = {21},
        pages = {21},
          doi = {10.1038/s43586-025-00392-0},
       adsurl = {https://ui.adsabs.harvard.edu/abs/2025NRvMP...5...21J}
}

@ARTICLE{Quintero2021,
       author = {{Quintero Noda}, C. and {Barklem}, P.~S. and {Gafeira}, R. and {Ruiz Cobo}, B. and {Collados}, M. and {Carlsson}, M. and {Mart{\'\i}nez Pillet}, V. and {Orozco Su{\'a}rez}, D. and {Uitenbroek}, H. and {Katsukawa}, Y.},
        title = "{Diagnostic capabilities of spectropolarimetric observations for understanding solar phenomena. I. Zeeman-sensitive photospheric lines}",
      journal = {Astronomy \& Astrophysics},
         year = 2021,
        month = aug,
       volume = {652},
          eid = {A161},
        pages = {A161},
          doi = {10.1051/0004-6361/202037735},
archivePrefix = {arXiv},
       eprint = {2106.05084},
 primaryClass = {astro-ph.SR},
       adsurl = {https://ui.adsabs.harvard.edu/abs/2021A&A...652A.161Q}
}

@ARTICLE{asplund,
       author = {{Asplund}, Martin and {Grevesse}, Nicolas and {Sauval}, A. Jacques and {Scott}, Pat},
        title = "{The Chemical Composition of the Sun}",
      journal = {Annual Review of Astronomy and Astrophysics},
         year = 2009,
        month = sep,
       volume = {47},
       number = {1},
        pages = {481-522},
          doi = {10.1146/annurev.astro.46.060407.145222},
archivePrefix = {arXiv},
       eprint = {0909.0948},
 primaryClass = {astro-ph.SR},
       adsurl = {https://ui.adsabs.harvard.edu/abs/2009ARA&A..47..481A}
}

@ARTICLE{Morton2023,
       author = {{Morton}, R.~J. and {Sharma}, R. and {Tajfirouze}, E. and {Miriyala}, H.},
        title = "{Alfv{\'e}nic waves in the inhomogeneous solar atmosphere}",
      journal = {Reviews of Modern Plasma Physics},
         year = 2023,
        month = dec,
       volume = {7},
       number = {1},
          eid = {17},
        pages = {17},
          doi = {10.1007/s41614-023-00118-3},
archivePrefix = {arXiv},
       eprint = {2208.05222},
 primaryClass = {astro-ph.SR},
       adsurl = {https://ui.adsabs.harvard.edu/abs/2023RvMPP...7...17M}
}

@ARTICLE{Yadav2021,
       author = {{Yadav}, N. and {Cameron}, R.~H. and {Solanki}, S.~K.},
        title = "{Slow magneto-acoustic waves in simulations of a solar plage region carry enough energy to heat the chromosphere}",
      journal = {Astronomy \& Astrophysics},
         year = 2021,
        month = aug,
       volume = {652},
          eid = {A43},
        pages = {A43},
          doi = {10.1051/0004-6361/202039908},
archivePrefix = {arXiv},
       eprint = {2105.02932},
 primaryClass = {astro-ph.SR},
       adsurl = {https://ui.adsabs.harvard.edu/abs/2021A&A...652A..43Y}
}



\end{document}